\documentclass[%
 reprint,
 amsmath,amssymb,
 aps,
]{revtex4-2}
\usepackage[caption = false]{subfig}
\usepackage{graphicx}
\usepackage{dcolumn}
\usepackage{bm}
\usepackage{hyperref}
\usepackage{xcolor}
\usepackage{amsmath,amsfonts,amssymb}
\usepackage[utf8]{inputenc} 
\usepackage[T1]{fontenc}
\def\apj{ApJ}
\def\apjl{ApJL}

\def\nat{Nature}

\def\aap{{A\&A}}
\def\prd{PRD}

\def\rmd{\textrm{d}}

\begin{document}
\title{Detecting Cosmic Strings with Lensed Fast Radio Bursts}
\author{Huangyu Xiao$^1$}
\author{Liang Dai$^2$}
\author{Matthew McQuinn$^3$}

\affiliation{$^1$Department of Physics, University of Washington,  Seattle, WA 98195,USA}
\affiliation{$^2$Department of Physics, 366 Physics North MC 7300, University of California, Berkeley, CA
94720, USA}
\affiliation{$^3$Department of Astronomy, University of Washington,  Seattle, WA 98195,USA}

\begin{abstract}

Correlated red noise recently reported from pulsar timing observations may be an indication of stochastic gravitational waves emitted by cosmic strings that formed during a primordial phase transition near the Grand Unification energy scale.
Unfortunately, known probes of cosmic strings, namely the Cosmic Microwave Background anisotropies and string lensing of extragalactic galaxies, are not sensitive enough for low string tensions of $G\mu = 10^{-10}-10^{-7}$ that are needed to explain this putative signal. 
We show that strong gravitational lensing of Fast Radio Bursts (FRBs) by cosmic strings is a potentially unambiguous avenue to probe that range of string tension values. The image pair of string lensing are expected to have identical magnification factor and parity, and have a typical time delay of $\sim 10^2\,\,(G\,\mu/10^{-8})^2$ seconds. The unique spectral fingerprint of each FRB, as well as the possibility to detect correlations in the time series of the electric field of the radio waves, will enable verification of the string lensing interpretation. Very-Long-Baseline Interferometry (VLBI) observations can spatially resolve the image pair and provide a lower bound on the string tension based on the image separation.
We calculate the FRB lensing rate as a function of FRB detection number for several different models of the FRB redshift distribution. We find that a survey detecting $\sim 10^5$ FRBs, in line with estimates for the detection rate of the forthcoming survey CHORD, can uncover a strong lensing event for a string tension of $G\mu \simeq 10^{-7}$. Larger FRB surveys, such as Phase 2 of the Square Kilometre Array (SKA), have the potential to significantly improve the sensitivity on the string tension to $G\mu \simeq 10^{-9}$.

\end{abstract}

\maketitle





\section{Introduction}



Cosmic strings are topological defects that form in the early Universe from first-order phase transitions that break either a global or a gauged $U(1)$ symmetry. These defects can survive until the present day \cite{Kibble:1976sj,Hindmarsh:1994re,Vilenkin:1984ib}. The energy scale of the phase transition might be close to that at the end of inflation \cite{Dvali:1998pa,Baumann:2007ah,Burgess:2001fx}, when the gauge symmetry associated with Grand Unification is broken \cite{Kibble:1982ae,Jeannerot:2003qv}, or when a global Peccei-Quinn symmetry is broken in the post-inflationary scenario \cite{Vilenkin:1982ks}. (The Peccei-Quinn strings can only survive till the late-time Universe if explicit breaking is small enough that the axion mass is smaller than the Hubble parameter.)
If cosmic strings do exist, they directly probe fundamental interactions in the early Universe at very high energy scales. Around the time of string formation, there is roughly one cosmic string per horizon scale. As the Universe expands, the strings subsequently undergo self or mutual intersections and form longer strings, so that a small number of strings remain per Hubble volume, with the typical strings being relatively straight on this scale \cite{PhysRevD.83.083514}. 

Several studies have suggested that the candidate stochastic gravitational wave background in the nanohertz frequency range recently reported from pulsar timing arrays might result from bursty gravitational wave emissions from a network of cosmic strings \cite{NANOGrav:2020bcs,Blasi:2020mfx,Ellis:2020ena,Buchmuller:2020lbh}. Studies have found that the tension $\mu$, which is defined as the energy per unit length, needs to be in the range of $G\mu\sim 10^{-10}$--$10^{-7}$, with the exact value required dependent on the loop size distribution of the string network (as string loops dominate the gravitational radiation \cite{Blasi:2020mfx}). Interestingly, this tension range corresponds to a symmetry breaking scale of $\sim M_{\rm pl}\sqrt{G\mu}= 10^{14}$--$10^{16}\rm GeV$ \cite{Hindmarsh:1994re,Bogomolny:1975de}, which is close to the energy scale of grand unified theories \cite{King:2020hyd} and the inflationary energy scale in models with a monomial inflaton potential \cite{Lyth:1998xn}.

The strongest limits on string tensions have come from analyses of the cosmic microwave background (CMB) anisotropies.
While cosmic strings had once been a leading candidate source for the primordial density fluctuations that seeded cosmic structure formation \cite{Turok:1984cn,Vilenkin:1983jv}, modern CMB observations have lent strong support to an early inflationary phase as the origin of structure formation \cite{2000cils.book.....L}. Nevertheless, cosmic strings may still have a small contribution to the density perturbations. Furthermore, cosmic strings can create discontinuities in the CMB temperature anisotropy via the Kaiser-Stebbins effect \cite{Kaiser:1984iv}.
The string tension is constrained by the CMB anisotropy to be $G\mu<1.1\times 10^{-7}$ at the $2\sigma$ level for ordinary Nambu-Goto string networks \cite{Charnock:2016nzm,Planck:2015fie}.

Another method for detecting cosmic strings is lensing of galaxies \cite{Sazhin:2006kf}. A background source exhibits two identical lensed images when it has a sufficiently small impact parameter relative to a foreground string. 
String lensing signatures have been sought for in the galaxy survey data \cite{Morganson_2010,Christiansen_2011,Gasparini:2007jj}. No convincing string lensing system has been found, despite of some candidate lenses reported in the past \citep{Sazhin:2006kf,Christiansen_2011,Morganson_2010}. Hence an upper limit on $G\mu < 3\times 10^{-7}$ has been placed from galaxy lensing \cite{Christiansen_2011}. Unfortunately, despite that modern photometric surveys have cataloged billions of galaxies, future surveys are unlikely to substantially tighten the constraint on the string tension. When the string tension is smaller than $G\mu\sim10^{-7}$, the angular separation between the two lensed images is smaller than $\sim1$ arcsecond if the lens and the source are at cosmological distances. This saturates the angular resolution limit of seeing-limited telescopes on the ground and those of the space telescopes are only up to an order of magnitude better. Furthermore, $0.2$~arcsecond is about the angular size of a $2$~kpc galaxy at $z=2$. Thus, imaging surveys of galaxies are unlikely to probe $G\mu \lesssim  3\times 10^{-8}$. 

Thus, CMB and galaxy lensing are not able to constrain tensions in the range of $G\mu\sim 10^{-10}$--$10^{-7}$ that are consistent with the putative gravitational wave background. In this work, we propose gravitationally lensed fast radio bursts (FRBs) as a probe of cosmic strings in this tension range. In contrast to galaxy lensing, where string lensing cosntraints are limited by angular resolution, lensed extragalactic FRBs can be unambiguously distinguished in the time domain.  This allows probing much lower tensions. We further show that there are a few avenues through which a string lensed FRB event can be further validated. 
Therefore, FRBs can open up a new window to detecting extremely localized, linear gravitating structures in the Universe and such a method can be potentially very powerful given the high FRB detection rates at CHORD and expected for the forthcoming Square Kilometre Array (SKA) and Deep Synoptic Array (DSA) \cite{2019clrp.2020...28V,2019BAAS...51g.255H}.\footnote{Relatedly, strong lensing of fast radio bursts (FRBs) has already been proposed as a powerful probe of massive compact halo objects such as primordial black holes in the mass range of $10^{-4}-10^4M_{\odot}$ \cite{Munoz:2016tmg,Kader:2022jqp}.} 


The remainder of this paper is organized as follows. In Sec.~\ref{sec:lensing}, we study the strong lensing effect caused by cosmic strings, which includes a calculation of the strong lensing rate for FRBs and a discussion of the lensing time delay. In Sec.~\ref{sec:radio}, we study how sensitive radio surveys must be in order to be able to detect an adequate number of FRBs. In Sec.~\ref{sec:discussion}, we discuss how to identify FRB lensing events as well as the unique features of string lensing. Concluding remarks will be made in Sec.~\ref{sec:concl}. Throughout this paper, we adopt a unit system in which the speed of light is set to be $c=1$.

\section{The strong lensing by Cosmic strings}
\label{sec:lensing}

\begin{figure}[h!]
\centering
\includegraphics[width=0.5\textwidth]{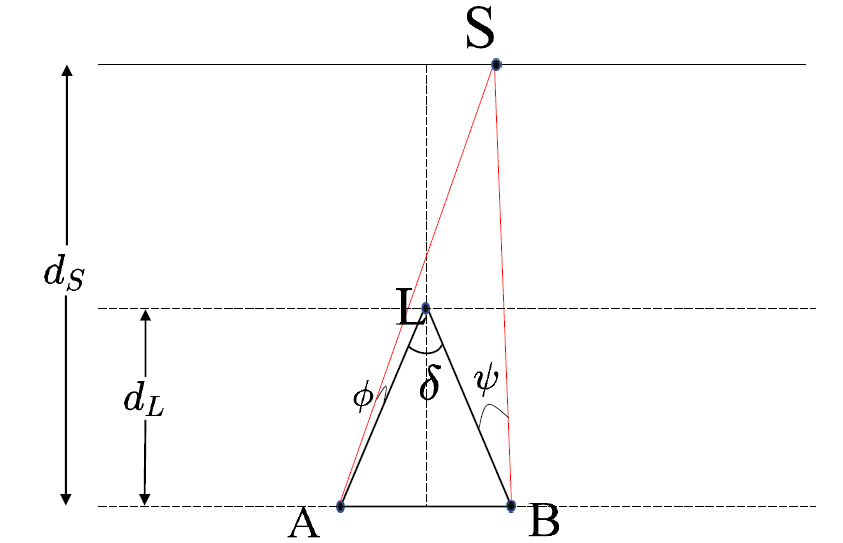}
\caption{Geometric illustration of strong gravitational lensing by a straight cosmic string. $\rm L$ marks the location of the string (perpendicular to the plane of the paper) while $\rm S$ marks the location of the point source. $\delta$ is the deficit angle. The two positions on both side of the conical cut, indicated as $A$ and $B$, represent the same physical location of the observer on Earth. The observer thus receives two rays coming at slightly different angles. The angular separation between the lens and the source, as observed from A and B, are denoted as $\phi$ and $\psi$, respectively. }
\label{fig:lens}
\end{figure}

Studies have found that the total length of a typical string network is primarily accounted for by string segments with large characteristic curvature radii \cite{Vachaspati:1984dz,Gorghetto:2018myk}, which are much larger than the angular scale of strong lensing we concern in this work. Therefore, we approximate strings as straight lines in calculations. It is well known that the spacetime metric around a straight massive string at rest is identical to that of the flat spacetime except that there is a conical deficit angle of $\delta=8\pi\,G\mu$ at the location of the string \cite{PhysRevD.23.852}. Consequently, the angular coordinate around the string runs from 0 to $2\pi-\delta$ and the geometry is that of a cone in the plane transverse to the string, as shown in Fig.~\ref{fig:lens}. The observer will see two lensed images of a source if the angular impact parameter is smaller than an Einstein angle $\theta_{\rm E}$. We define this situation where two lensed images form as the string strong lensing regime.

Since the space is locally flat, there is no local distortion to photon propagation. The deficit angle $\delta=8\pi G\mu$ characterizes the lensing properties of a cosmic string and fully specifies the space-time geometry around it. Given the deficit angle $\delta$, as well as the source and lens distances, the Einstein angle is given by $\theta_{\rm E}$ \cite{1984ApJ...282L..51V, Sazhin:2006kf}:
\begin{equation}
    \theta_{\rm E}=4\,\pi \, G\mu\, {\rm sin\,}i\,\frac{d_{\rm LS}}{d_{\rm S}},
\end{equation}
With the notation for the comoving distance $D_c({z_1,z_2})$ from one redshift $z_1$ to another redshift $z_2>z_1$, we define the various angular diameter distances: $d_{\rm LS}= (1+z_S)^{-1}\,D_c(z_L, z_S)$ is the angular diameter distance from the lens to the source, $d_{\rm S}=(1+z_S)^{-1}\,D_c(0, z_S)$ is the angular diameter distance from the Earth to the source. The geometric factor ${\rm sin} \, i$ accounts for the projection of the cosmic string in the plane of the sky, where $i$ is the inclination, i.e. the angle the string makes with respect to the line of sight. 

For string tension values that are of interest in this work, $G\mu \sim 10^{-8}$, the angular separation between the two lensed images is $2\,\theta_E \sim 4\pi \times 10^{-8} {\rm ~~ radian}=0.025^{\prime\prime}$, for a string that lies in the plane of the sky and is located halfway between the Earth and the source.
When observing at 1 GHz, such small angular separations are only resolvable with a baseline of a thousand kilometer long, which requires the technique of Very-Long-Baseline Interferometry (VLBI). If such a lensing event is confirmed by VLBI, the angular separation $2\,\theta_{\rm E}$ will be measurable and we will obtain a lower bound on the dimensionless string tension $G\mu$ since $({\rm sin\,}i)\,(d_{\rm LS}/d_{\rm S})<1$. This will be an important piece of information to guide other string searches.
However, even without spatially resolving the two lensed images, we will still be able to identify lensed FRB events because of the time delay between the two images. 

Understanding the lensing effect of a single string, we now study the total number of strings in the Universe in order to calculate the strong lensing rate. Numerical studies have found that the number of cosmic strings per horizon volume grows logarithmically with time \cite{Almeida:2021ihc,Buschmann:2021sdq,Gorghetto:2018myk}. The proper number density of strings can be parameterized in the following form \cite{Buschmann:2021sdq}
\begin{equation}\label{eq:scaling_re}
n_{\rm str}(z)
    =\overbrace{\left(\alpha_1+\alpha_2\,{\rm ln}\left[\frac{f}{H(z)} \right] \right)}^{\xi}\,H(z)^3.
\end{equation}
Ref.~\cite{Buschmann:2021sdq} have derived $\alpha_1=-1.82$ and $\alpha_2=0.254$ from axion string simulations, $f$ is the temperature of the phase transition that leads to string formation, and $H(z)$ is the Hubble parameter as a function of the redshift \footnote{Axion strings arise from a broken global symmetry and hence have a more extended energy density profile than gauge strings do, which may affect the above scaling.}.
If the phase transition temperature is $\sim 10^{15}\rm GeV$
, we expect to have $\sim30$ strings within the present-day Hubble volume. The scaling relation Eq.~(\ref{eq:scaling_re}) is likely to apply to other string models such as Nambu-Goto strings with the number of strings per horizon on the same order of magnitude \cite{Allen:1990tv,Blanco-Pillado:2011egf}.

\subsection{Lensing rate of FRBs}
Consider a straight string segment of proper length $L_{\rm str}$, inclination $i$, and at lens redshift $z_L$, the ``super-critical'' area cast on the source plane where a source is lensed into two images is
\begin{equation}
    \sigma = \Tilde{L}_{\rm str}\,\Tilde{d} =\frac{d_S}{d_L}\,(L_{\rm str}  \, {\rm sin\,}i) (2\,\theta_E)\,d_{\rm S},
\end{equation}
where $\Tilde{d}=2\,\theta_E\,d_{\rm S}$ is the width of the lensed area on the source plane while $\Tilde{L}_{\rm str}=(d_S/d_L)\,L_{\rm str}  \, {\rm sin\,}i $ is the projected string length on the source plane.
Summing over string segments of all orientations within the volume element gives the total ``super-critical'' area cast on the source plane
\begin{equation}
    \sigma_{\rm tot} = \langle\, {\rm sin^2\,}i\rangle_{\Omega}\,\int \rmd V\,\frac{\rmd {\cal L}_{\rm tot}}{\rmd V}\,(8\pi\,G\,\mu)\,d_{\rm LS}\,\frac{d_S}{d_L}, 
\end{equation}
where $\langle\, {\rm sin^2\,}i\rangle_{\Omega}=\int\rmd i \, {\sin^2}i \,{\cos}i=2/3$ results from averaging over string orientations.
The total string length per volume can be expressed as 
\begin{equation}
    \frac{\rmd{\cal L}_{\rm tot}}{\rmd V} = \langle L_{\rm str} \rangle \, n_{\rm str} =  \xi\,H(z_L)^2,
\end{equation}
where we have made the approximation that the average string length is $\langle L_{\rm str} \rangle = 1/H$ as $\sim 80\%$ of the strings have this length \cite{Gorghetto:2018myk}. 
The probability for a FRB at $z_S$ to be strongly lensed is 
\begin{equation}
        P(z_S) 
        =\frac{16}{3}\pi\,G\mu \int_0^{z_S}\rmd z_{L}\,\xi\,\frac{d_{\rm L}d_{\rm LS} H(z_L)}{d_{\rm S}}\,\frac{1}{1+z_L}.
\end{equation}
The integrated probability of observing a lensed FRB, given the FRB redshift distribution, can be calculated as
\begin{equation}
    P_{\rm obs}= \int \frac{\rmd^2 N_{\rm FRB}}{\rmd \Omega\,\rmd z_S}\,P(z_S)\,\rmd \Omega\,\rmd z_S.
\end{equation}

\subsection{Redshift Distribution of FRBs}

The FRB redshift distribution needs to be known for predicting the cosmic string lensing rate.  Precise knowledge about how the intrinsic FRB rate evolves with cosmic time is lacking, and the observed rate is sensitive to the FRB luminosity function, which is not yet well constrained.  We start with two different models for the intrinsic FRB redshift distribution, with one of them tracking the star formation rate, and the other tracking the accumulated stellar mass. While the former model appears to be more consistent with the young magnetar origin for FRBs \citep{CHIMEFRB:2020abu, 2020Natur.587...59B}, it has been suggested that there may be a sizeable FRB sub-population that does not trace the star formation rate. In fact, a couple FRBs have been localized to galaxies without significant star formation \citep{Li:2020esc}, and one even to a globular cluster \citep{2021ApJ...910L..18B,Kirsten:2021llv}. 

The CHIME FRB survey has found a power-law cumulative distribution for the fluence in the $400-800$ MHz frequency range,
\begin{equation}
    N_{\rm FRB}(>F)\propto F^{\alpha},
\end{equation}
with $\alpha=-1.4$. Here, we define the fluence of an FRB, $F$, as the flux multiplied by burst duration. By modeling the FRB redshift distribution, we aim to reproduce the fluence power-law distribution in the range where observational data are available.
In principle, neither fluence nor flux is the precise quantity that sets the detection threshold. Flux multiplied by the number of independent temporal samples should be the most appropriate quantity. However, many FRBs are temporally unresolved by the detection instrument even at the millisecond temporal resolution. For these FRBs, fluence can be directly related to the detection $S/N$. 
\begin{figure}[h]
\includegraphics[width=8cm]{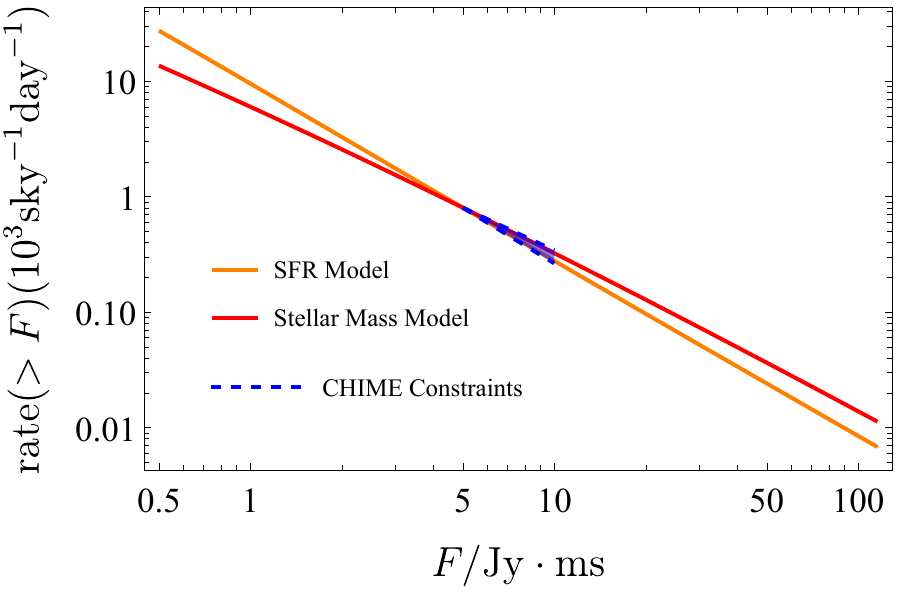}
\caption{FRB fluence distribution predicted by our adopted models compared to that inferred from CHIME data. Our two model fluence distributions lie within the band whose width is determined by observational uncertainties. This shows that our adopted models are compatible with CHIME survey results. }
\label{fig:fluence_comparison}
\end{figure}

\begin{figure}[h]
\includegraphics[width=8cm]{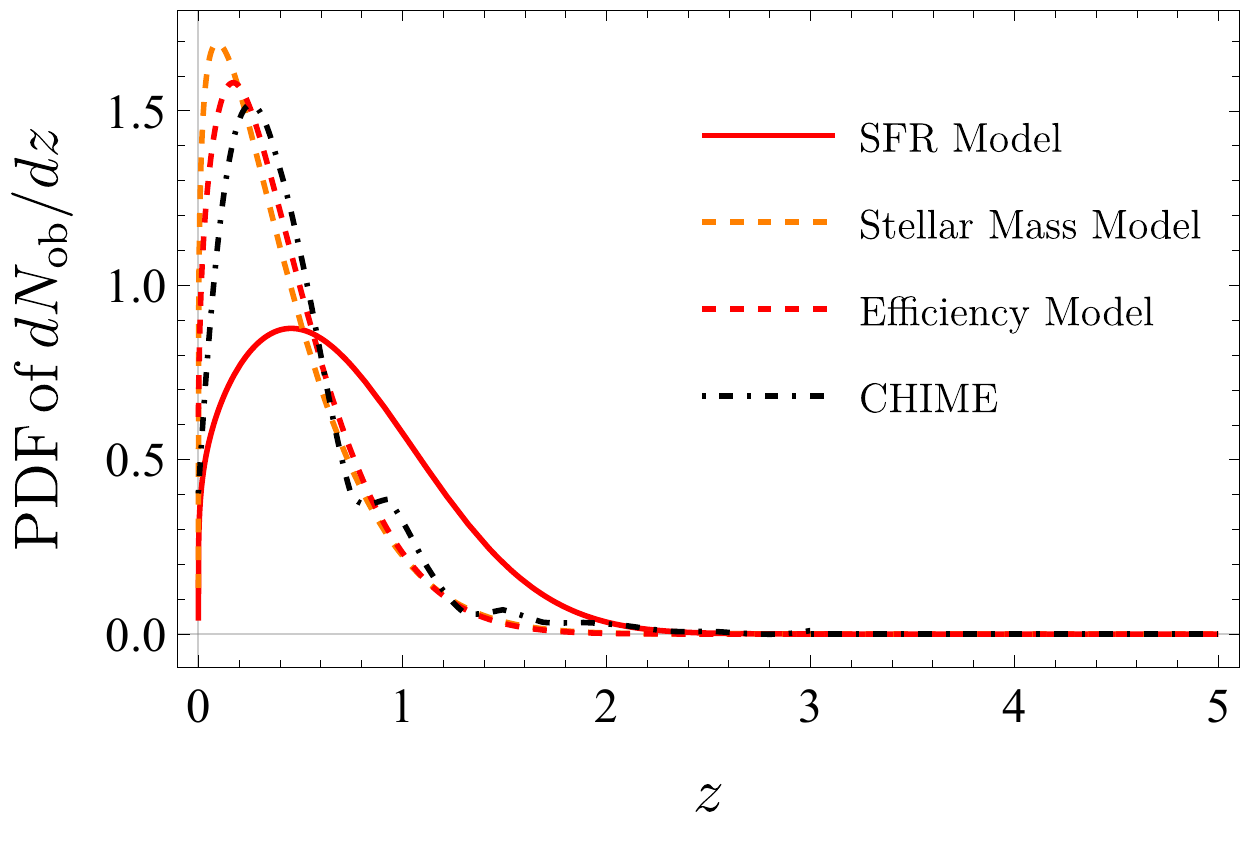}
\caption{Probability distribution function (PDF) of the observed FRB redshift distribution for a number of models. This PDF differs from the intrinsic redshift distribution as fainter ones at higher redshifts are undetectable. Therefore, the detection threshold of the FRB survey instrument determines the observed $z$ distribution. 
We plot for the Star Formation Rate model (red solid curves) with a fluence threshold $F_{\rm min}=0.4\rm Jy\cdot ms$. The Stellar Mass model also uses a fluence threshold $F_{\rm min}=0.4\rm Jy\cdot ms$.
We also plot a efficiency model which assumes less detection efficiency for low fluence FRBs and it assumes the star formation model for the intrinsic distribution. 
The black dashed curve shows the redshift distribution of FRBs observed by CHIME under the assumption that the FRB host galaxy has a negligible contribution to the dispersion measure so that the dispersion measure can be mapped directly to a redshift.}
\label{fig:pdf}
\end{figure}
The intrinsic FRB luminosity can be observationally well fitted by a Schechter function \cite{Luo:2018tiy,Luo:2020wfx} --- a power law followed by an exponential cutoff at the high luminosity end \cite{1976ApJ...203..297S}:  
${\rmd \dot n_{\rm FRB}(z)}/{\rmd E} \propto \,E^{-\beta}\,e^{-E/E_0}$,
with a power-law index $1.8\lesssim\beta\lesssim 2$. We combine an intrinsic redshift distribution with the intrinsic luminosity function to reproduce a fluence distribution compatible with CHIME observations.
We assume an intrinsic redshift distribution 
\begin{equation}
     \frac{\rmd \dot n_{\rm FRB}(z)}{\rmd E} \propto \,E^{-1.9}\,e^{-E/E_0}\,{\rm f}(z),
\end{equation}
where ${\rm f}(z)$ shall be determined by the specific models of the intrinsic redshift distribution.
We set a burst energy cutoff $E_0= 1.5\times 10^{41}\rm erg$ as this allows us to fit the CHIME redshift distribution with a model where FRBs trace the star formation rate (see below); however, others have adopted an energy cutoff $E_0= 3\times 10^{41}\rm erg$  \cite{Luo:2020wfx,Zhang:2020ass,Zhang:2021kdu}, which has the effect of pushing the observed FRB distribution to higher redshifts (and would strengthen our constraints). 
The comoving star formation rate density, $\rho_{\rm SFR}(z)$, is well constrained by observations. It can be approximated by the following formula \cite{Qiang:2021ljr}
\begin{equation}
    {\rm f}(z)=\rho_{\rm SFR}(z)\propto\frac{(1+z)^{2.6}}{1+[(1+z)/{3.2}]^{6.2}}.
\end{equation}
As $\rho_{\rm SFR}(z)$ peaks at $z=2-3$, in this model a majority of the FRBs come from this redshift range. Alternatively, we may assume that the FRB rate tracks the integrated star formation over time, which leads to the Stellar Mass model, for which 
\begin{equation}
  {\rm f}(z)= \int_z^{\infty}\,\rmd z^{\prime}\,\frac{\rho_{\rm SFR}(z^{\prime})}{(1+z^{\prime})\,H(z^{\prime})}.   
\end{equation}
The redshift distribution of the observed FRBs depends on the sensitivity of the radio telescope. Hence, we introduce a fluence threshold above which an FRB is detectable to the telescope. Integrating the luminosity function above the detection threshold and over the entire range of redshifts, we obtain the observed redshift distribution of FRBs brighter than a given fluence: 
\begin{equation}\label{eq:FZ_dis}
\begin{split}
        \frac{\rmd N_{\rm FRB}(z,>F)}{\rmd z}=&  \int_{E_{\rm min}(F)}^\infty \rmd E\,  \overbrace{(1+z)^{-1}}^{\rm time~dilation}\\
        &\times\frac{\rmd \dot n_{\rm FRB}(z)}{\rmd E}\,\left(4\pi D_c^2 \right)\,\frac{\rmd D_c}{\rmd z},
\end{split}
\end{equation}
where $E_{\rm min}(F) = F\,(4\pi\,d_L^2)\,K_{\nu}/(1+z)$ is the relation between the luminosity and the fluence in the band the telescope observes, 
$d_L$ is the luminosity distance and $K_{\nu}=(1+z)^{-s}$ accounts for the frequency spectral index, whose average value is found to be $s= -1.5^{+ 0.2}_{- 0.3}$ \cite{2019}; we use $s = -1.5$ for our calculations, but the results are only weakly sensitive to this choice. $K_{\nu}$ appears in the expression because the flux and luminosity to be measured are not bolometric but densities per unit frequency and an observed FRB at redshift $z$ with observed frequency $\nu_o$ has an original frequency of $\nu_e=(1+z)\nu_o$. 

Integrating Eq.~(\ref{eq:FZ_dis}) over redshift, we obtain the fluence distribution, with a normalization we empirically fix based on the CHIME FRB catalog. This is shown in Fig.~\ref{fig:fluence_comparison}. The CHIME data constrains the FRB fluence distribution to follow a power law in the range $5$--$10\,{\rm Jy}\cdot {\rm ms}$, which is marked as the narrow blue band in Fig.~\ref{fig:fluence_comparison}. This shows that we can reproduce a power-law fluence distribution consistent with data. 

By selecting an appropriate fluence threshold 
, we can predict the observed redshift distribution with Eq.~(\ref{eq:FZ_dis}).
As shown in Fig.~\ref{fig:pdf}, the FRB redshift distribution peaks at $z\sim 1.8$ if the fluence threshold is $F_{\rm min}=0.4\,\rm Jy\cdot ms$. In addition to the redshift distibution models, Fig.~\ref{fig:pdf} shows the redshift distribution of CHIME FRBs \emph{if} the dispersion measure (DM) is of intergalactic origin and not intrinsic to the host. Under this assumption, the DM can be converted into the source redshift, except that the true redshift can be overestimated if there is a sizeable host DM contribution. The obtained redshift distribution appears to agree better with the Stellar Mass model rather than the SFR model.
CHIME and other radio telescopes can detect brighter bursts in the outskirts of the beam. This results in an effective survey area that is larger for brighter (and typically lower redshift) FRBs.  
We adopt the model of CHIME detection efficiency in Ref.\cite{Zhang:2021kdu} to correct for this effect. By making this correction, we confirm that we fit well the fluence distribution of the CHIME FRB catalog.  
This efficiency model allows us to understand how our FRB redshift distributions may be influenced by detector efficiency. This correction for the $0.4$ Jy$\cdot$ ms star formation model is shown in Figure~\ref{fig:pdf}, acting to shift the observed redshift distribution to lower redshifts. After adopting the efficiency model, the redshift distribution of star formation model can match CHIME well, which suggests the FRB may track the star formation rate. 
Other FRBs surveys may suffer less of an efficiency correction as CHIME and should be sensitive to higher redshift bursts: FRBs detected with Parkes show on average higher DM than CHIME \citep{Petroff:2016tcr}.



Now that we have built a model for the FRB redshift distribution, we are ready to compute the probability of lensing as a function of string tension. It is worth noting that the redshift distribution depends on the fluence threshold of detection, which varies with the collecting area of the surveys, as we will discuss later in Eq.~\ref{eq:Fmin}. Future radio surveys will have improved sensitivity thus lower fluence threshold than CHIME.  
We assume the fluence threshold is $F_{\rm min}=0.04\,\rm Jy\cdot ms$ 
for detection and study both the SFR model and the Stellar Mass model. This gives us the following strong lensing probability given the number of observed FRBs, $N_{\rm FRB}$: 
\begin{equation}\label{eq:detection_rate}
    P \approx \left(\frac{N_{\rm FRB}}{10^5}\right)\left(\frac{N_{\rm str}}{30}\right)
    \begin{cases}
       (G\mu) /(1.9\times10^{-7}) & \text{SFR; } \\
      (G\mu) /(5.2\times10^{-7}) & \text{Stellar Mass. }
    \end{cases}
\end{equation}
If distances to the FRBs are on average larger, we have a higher chance of detecting string strong lensing events. When the detection threshold is as low as $F_{\rm min} = 10^{-3}\,\rm Jy\cdot ms$, which is possible for future surveys with a larger collecting area, the detection probability can be further increased by about a factor of two.

\subsection{Time Delay}

Radio waves associated with the two lensed images travel along slightly different paths, and hence differ in the arrival time. In the picture where the space geometry around a long straight massive string has a deficit angle $\delta=8\pi G\mu$ but is otherwise a flat one (as shown in Fig.~\ref{fig:lens}), this time delay can be interpreted as a purely geometrical path length difference rather than a Shapiro time delay \cite{1985ApJ...288..422G}. This geometrical time delay is given by
\begin{equation}
       \Delta t = \frac{1}{2}\,\left( 1+z_{\rm L}  \right)\,d_L\,\Delta\theta\,|\phi-\psi|,
\end{equation}
where $\Delta\theta=\delta \, {\rm sin\,}i$ and $i$ is the angle between the string and the line-of-sight, $d_L$ is the angular diameter distance to the string lens, and the geometrical meanings of the angles $\phi$ and $\psi$ are shown in Fig.~\ref{fig:lens}. The lens equation can be expressed as $\phi+\psi=\Delta\theta\,(1-d_L/d_S)$.
For a typical example, if we set $d_L=d_S/2$ and $\phi=\Delta\theta/6$, we have $\psi=\Delta\theta/3$ and $\Delta t=d_L\,\Delta\theta^2/12$, for which the time delay is estimated to be:
\begin{equation}
\label{eq:timedelay}
\Delta t\sim 500\,{\rm s}\,\left(\frac{G\mu}{10^{-8}}\right)^2\left(\frac{D}{\rm Gpc}\right),
\end{equation}
where $D$ is the characteristic distance scale involved in the lens-source configuration.
For $G\,\mu \gtrsim 10^{-10}$, the two lensed images have a typical time delay that is larger than the FRB temporal width, and hence can be clearly separated. Here, we have assumed that both lensed images are observed by the radio telescope, which is not necessarily the case for $G\mu\sim 10^{-7}$ as the time delay is too long and the telescope might have a different pointing when the second image arrives. If the telescope direction rotates with the Earth (as is the case for a transit telescope), both images are detected if the Earth's rotation does not take the second image out of the beam, i.e. 
\begin{equation}
\Omega_{\rm FOV}\gtrsim 10^{-3}(D/{\rm Gpc})(G\mu/10^{-8})^2 \cos^2 b,
\end{equation}
where $b$ is the source's declination. This can pose a challenge for detecting string lensing events with a tension as large as $10^{-7}$ but can be fixed if some other telescopes follow up with the same pointing after each burst.   A similar problem (and solution) exists for multi-science instruments like the SKA that are targeting various sources.
In \S~\ref{sec:discussion}, we shall discuss how to distinguish lensed FRBs from repeating FRBs.

Interestingly, in the lucky case of a repeating FRB source being strongly lensed by a cosmic string, the transverse velocity and tension of the string can be constrained. Because a cosmic string is typically moving at a moderately relativistic speed $v_s \sim 0.1\,c$ \cite{Hindmarsh:1994re,Agrawal:2020euj}, different bursts from the same FRB source have different impact parameters with respect to the string, and hence have slightly different time delays. This effect has been studied for FRBs with multiple images lensed by other forms of gravitational lenses \cite{Dai:2017twh, Pearson2021lensedFRBgw, Wucknitz2021lensedFRBcosmology}. 
The change in the angular impact parameter between repeated FRB bursts is $\delta \phi=-\delta\psi=\delta d_t \,{\rm sin}k/d_S$, where $k$ is the inclination of the string velocity vector and $\delta d_t$ is the angular diameter distance traveled by the string between the bursts due to the transverse velocity. 
The time delay variation is given by 
\begin{equation}
    \delta t= (1+z_{\rm L})\,\delta d_t\,\Delta\theta\,(d_L/d_S).
\end{equation}
If the distances from the lens and the source are on the same order of magnitude, the time delay variation has a typical size
\begin{equation}
\label{eq:deltat}
    \delta t\sim  0.8\,{\rm s}\left(\frac{ v_s \,{\rm sin}k}{ 0.1 c}\right)\left(\frac{T_{\rm obs}}{\rm 1 yr}\right)\left(\frac{G\mu}{10^{-8}}\right),
\end{equation}
where $v_s$ is the string speed and $T_{\rm obs}$ is the observation time span. Therefore, lensed repeating FRBs will enable us to constrain the lens transverse velocity. Given existing constraints that $G \mu$ is unlikely to be larger than $10^{-7}$, the inferred mildly relativistic string speed, much higher than those of the other lens types, will be a smoking gun. 

It is worth noting that the angular separation of the two lensed images is always $2\,\theta_E$ independent of $\phi$ and $\psi$. Numerically, the image separation is on the order $\Delta\theta = 2\,\theta_E\sim 8\pi \times 10^{-8}\,{\rm rad}=0.02''$ for $G\mu\sim 10^{-8}$, which is spatially resolvable with radio VLBI observations. With such observations the host galaxy of the source is likely to be identified, but not for the string itself. Then it will be possible to derive a lower bound on $G\mu$ from the image separation $\Delta \theta=2\,\theta_E$ since $\theta_E<4\,\pi \, G\mu$. Measuring the time delay, which is given by $\Delta t =g\,\left( 1+z_{\rm L}  \right)\,d_L\Delta\theta\,\theta_E$ where $g= |\phi-\psi|/|\phi+\psi|<1$ is a geometric factor determined by angles, will thus lead to a lower bound on the distance to the string. In practice, this lower bound is likely to lie within a factor of two of the true value since it is unlikely that $g$ has a value that is very close to zero. It is therefore likely that the distance to the string lens can be inferred with a factor-of-two uncertainty. The ratio of the time delay change to the time delay is $\delta t/\Delta t = \delta d_t/(g\,\theta_E\,d_S)$, which will place an upper bound on the transverse speed of the lens since both $\theta_E$ and $d_S$ will be known. Again, it is likely that this upper limit is within a factor of two of the truth, so that one will be able to derive the string speed from the time delay variation up to uncertainty about the value of the geometric actor $g$, a mildly relativistic speed will be the smoking gun for a cosmic string.



\section{Forecasts for FRB detection rates}\label{sec:radio}
Here we discuss what $G\mu$ can be constrained by FRB surveys through a search for lensed events. The SNR of an FRB detection is given by the radiometer equation:
\begin{equation}
\frac{S}{N}=\frac{A\,F}{2\,k_{\rm B} T_{\rm sys}\tau_0}\sqrt{\Delta\nu\, \tau_0},
\end{equation}
where $F$ is the FRB fluence, $\tau_0$ is the duration of FRBs, $A$ is the total collecting area of the receiving dish(es)
, $T_{\rm sys}$ is the system temperature which is determined by the sky or instrumental background, and $\Delta \nu$ is the observation bandwidth. Numerically, the threshold fluence for detection is 
\begin{equation}\label{eq:Fmin}
\begin{split}
        F_{\rm min}&={4.4\rm ~Jy\cdot ms}\,\left(\frac{1000\rm \, m^2}{A}\right)\,\left(\frac{\Delta\nu}{\rm 100\,MHz}\right)^{-\frac12}\,\\
        &\times \left(\frac{\tau}{\rm 1\, ms}\right)^{-1/2}\left(\frac{S/N}{10}\right)\,\left( \frac{T_{\rm sys}}{50\,{\rm K}}\,\right).
\end{split}
\end{equation}

\begin{figure}[h!]
\includegraphics[width=8cm]{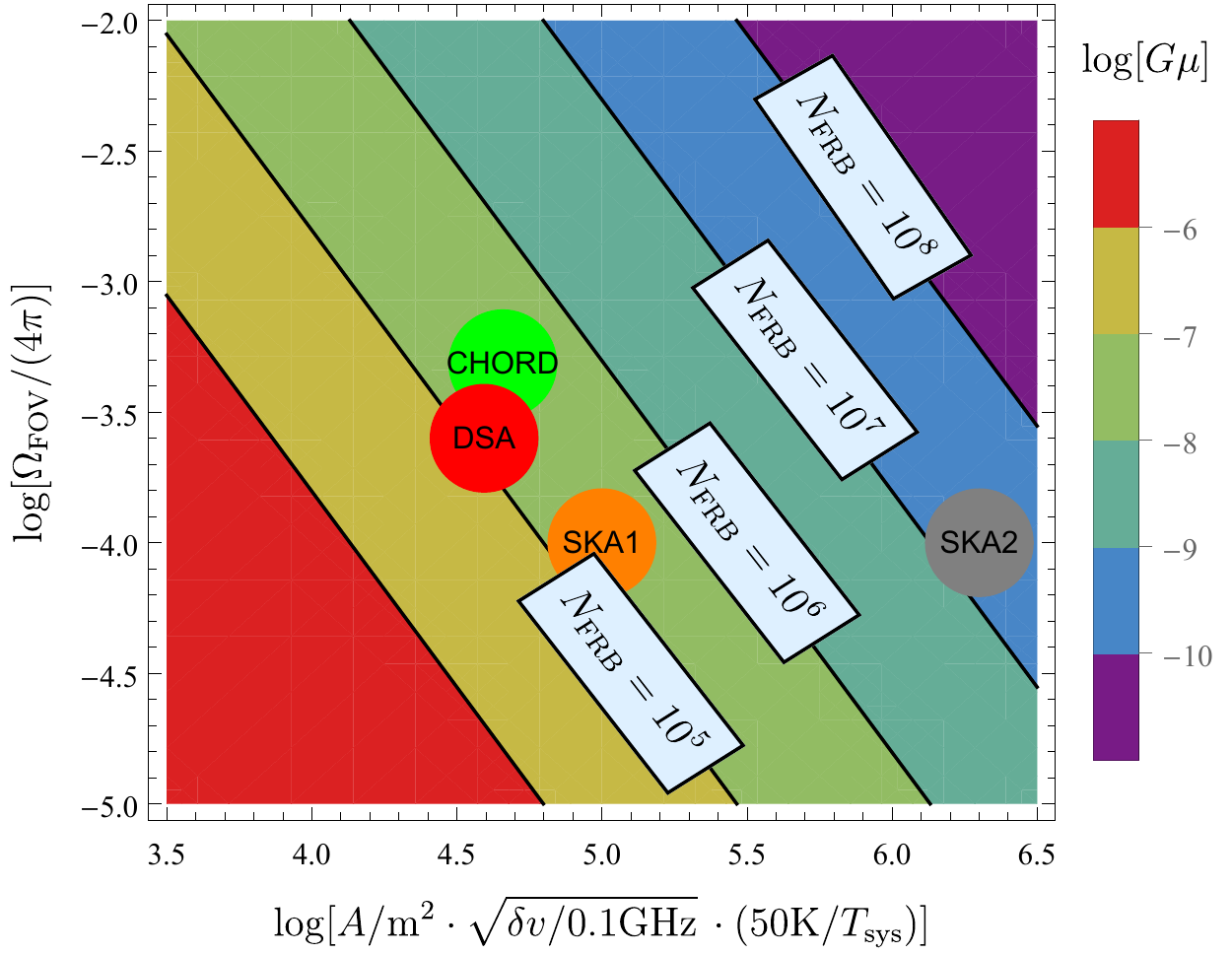}
\caption{
Forecast of sensitivity on the cosmic string tension $G \mu$ as a function of the survey sky coverage $\Omega_{\rm FOV}/4\pi$, the total radio telescope collecting area $A$, and the total number of surveyed FRBs detected $N_{\rm FRB}$. A ten-year survey is assumed, and the sensitivities that correspond to CHORD, DSA-2000, SKA Phase 1 and SKA Phase 2 are marked. We project that SKA Phase 2 will detect $N_{\rm FRB} \sim 10^7$ FRBs in the future and probe string tensions as small as $G\mu \sim 10^{-9}$.}
\label{fig:sensitivity}
\end{figure}

Having derived the fluence threshold for detection, we can calculate the FRB detection rate. 
The detection rate above a fluence threshold is expected to scale as $R(>F_{\rm min})\propto F^{-3/2}_{\rm min}$ if FRB sources are uniformly distributed in a Euclidean space \cite{Petroff:2019tty}. Accounting for the fact that the expanding Universe is not truly spatially Euclidean, the power law index is only somewhat altered for $\Lambda$CDM (but is dependent on the source luminosity function unlike the Euclidean limit). If we use the Euclidean scaling law as a good approximation, and take into account that CHIME detected $\sim 820$ FRBs per day per sky above 5 $\rm Jy\cdot ms$ \cite{CHIMEFRB:2021srp}, the FRB rate is
\begin{equation}
\begin{split}
        R&\sim 820\,{\rm day^{-1}}\,\left(\frac{\Omega_{\rm FOV}}{4\pi}\right)\,\left(\frac{F_{\rm min}}{5\,\rm Jy\cdot\rm ms}\right)^{-\frac32}\\
        &\sim 990 \,{\rm day^{-1}}\,\left(\frac{\Omega_{\rm FOV}}{4\pi}\right)\,\left(\frac{A}{10^3\,\rm m^2}\right)^{\frac32}\,\left(\frac{\Delta\nu}{\rm 0.1\,GHz}\right)^{\frac34}\,\\
        &\left(\frac{\tau}{\rm 1\, ms}\right)^{\frac34}\left(\frac{S/N}{10}\right)^{-3/2},
\end{split}
\end{equation}
where $\Omega_{\rm FOV}$ is the instantaneous sky coverage of the telescope beam, i.e. the field of view at one time.

\begin{figure}[h!]
\includegraphics[width=0.5\textwidth]{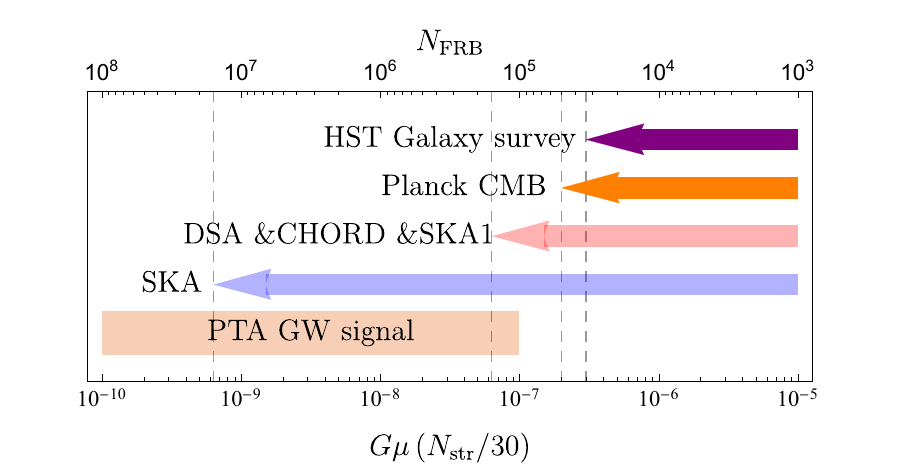}
\caption{Sensitivity for string tension $G\mu$ with different radio surveys assuming the star formation model. The peach bar indicates the string tension range that can be consistent with the candidate stochastic gravitational signal detected by pulsar timing arrays \cite{NANOGrav:2020bcs,Blasi:2020mfx}. The pink and blue arrow plots the potential sensitivity of DSA/CHORD/SKA1 and the future SKA2, using our estimates for the number of FRBs shown in Fig.~\ref{fig:sensitivity}. When $G\mu$ is as large as $\sim 10^{-7}$, dedicated follow-up observations will be needed; otherwise the second image might be out of the field of view due to a long time delay.
If no cosmic strings are detected, the SKA1 constraint will have an upper bound near the tip of the blue arrow. Orange and purple bars indicate the current constraints from Planck CMB \cite{Charnock:2016nzm,Planck:2015fie} and Hubble Space Telescope (HST) galaxy surveys \cite{Morganson_2010,Christiansen_2011}, respectively. 
}
\label{fig:result}
\end{figure}

Assuming ${\rm SNR}=10$, a characteristic duration of 1 ms for the FRB bursts and a system tempearture of 50 K (25 K for SKA), we can calculate the detection rate as a function of only two instrument parameters, the collecting area $A$ and the field of view $\Omega_{\rm FOV}$. Accounting for the number of FRB detections and the number of cosmic strings within the cosmic horizon, we obtain our forecast for string lensed FRBs (Eq.~(\ref{eq:detection_rate})). The FRB detection rate we use is consistent with previous predictions of $>10^4$ per year for CHORD and DSA-2000 \cite{2019clrp.2020...28V,2019BAAS...51g.255H}. In general, radio surveys with a large collecting area $A$, a large field of view $\Omega_{\rm FOV}$ and a large bandwidth are advantageous for detecting more FRBs, and hence more sensitivity to cosmic strings.
Assuming a survey that lasts for 10 years and $50\%$ of the time is actively taking data for FRBs, we derive the sensitivity to the string tension $G \mu$ for different sets of observation parameters that correspond to different forthcoming radio surveys. This is shown in Fig.~\ref{fig:sensitivity}. Future radio surveys are expected to detect as many as $N_{\rm FRB} \sim 10^6-10^7$ FRBs, making them a very powerful probe of cosmic strings. SKA Phase 2 will provide the best sensitivity owing to its large collecting area, while CHORD and DSA are also expected to detect a large number of FRBs. Bustling Universe Radio Survey Telescope for Taiwan (BURSTT) \cite{Lin:2022wgp} will also detect $\sim 10^4$ FRBs per year owing to its large field of view and Packed Ultra-wideband Mapping Array (PUMA) \cite{PUMA:2019jwd} can even reach a similar sensitivity to SKA-2.
It is worth noting that none of these existing surveys are ideal for detecting as many FRBs as possible since the field of view is not wide enough. A large number of small dishes or dipoles that can achieve a large field of view and a large collecting area simultaneously can maximize the number of detection and thus the sensitivity to cosmic strings.
Finally, we compare the sensitivity forecast from FRB observations to the current constraints from galaxy surveys and CMB in Fig.~\ref{fig:result}. A large fraction of the string tension range compatible with the candidate stochastic gravitational wave background detected by pulsar timing \cite{NANOGrav:2020bcs,Blasi:2020mfx} can be probed by FRB string lensing. 

\section{Discussion}
\label{sec:discussion}

In this section, we discuss how to distinguish the strong lensing event from two different FRB sources and how to distinguish lensing by cosmic strings from that by other lenses (e.g. galaxies).
FRBs have unique fingerprints on the spectrum, which can tell if this is an lensing event. The electric fields of two images can also be correlated, which we will discuss in more detail on the possible decoherence. 
The unique feature of string lensing is that lensed images are expected to have exactly the same magnification up to the small weak lensing effect. We will estimate the magnification difference caused by weak lensing.
Also, strings have mildly relativistic speeds $\sim 0.1\,c$, which far exceed those of other astrophysical lenses. The high speed can be measured if the lensed FRB repeats. 

\subsection{Spectra and electric field correlations}
We expect that two physically distinct FRB bursts will show different features in the frequency spectrum, while two lensed images of the same FRB source will manifest the same spectrum. An FRB frequency spectrum often exhibits a sufficiently complex pattern which serves as a unique fingerprint to distinguish different bursts \citep{Macquart_2019}.  

A even more foolproof test of lensing would be to correlate the electric field time series of the two bursts to confirm that they are lensed images. The conditions for the electric field time series to correlate are first that the FRB emitting region $\ell_{FRB}$ is unresolved by the string lens, requiring
\begin{equation}
    \ell_{\rm FRB} \lesssim \frac{\lambda}{\theta_E d_{\rm L}} d_{\rm LS} \sim 10^8 {\rm cm} ~ \left( \frac{\lambda}{10\, {\rm cm}} \right)  \left(\frac{G\mu}{10^{-8}}\right)^{-1} \left(\frac{d_{\rm LS}}{d_{\rm L}} \right).
\end{equation}
This condition is satisfied for the $G\mu\sim 10^{-9}-10^{-7}$ of interest for emission on a scale comparable to the neutron star size of $\sim 10~$km, which occurs in magnetospheric models for FRB emission. However, the condition may not be satisfied in external shock models for FRB emission, which predict $\ell_{FRB} \sim 10^9-10^{11}$cm after the relativistic aberration effect is accounted for \cite{Metzger:2019una}\footnote{In detail, $\ell <  (2 \Gamma^2 \, c \, \delta t_{\rm var})/\Gamma$, larger by a factor of the Lorentz factor but still smaller than we require for $\Gamma \lesssim 10^3$ for a variability timescale of $\delta t_{\rm var} = 1$ ms, a condition on the Lorentz factor that is satisfied for the fiducial parameters taken in external shock models.  Some bursts suggests $\delta t_{\rm var} \lesssim 1 ~\mu$s.}.
Additionally, some FRBs are observed to be scattered by the host system, which would result in a larger effective $\ell_{FRB}$ of $ 3\times 10^{8} (d_{\rm screen}/0.01\; {\rm pc})^{1/2} (\tau_{\rm sc}(1~{\rm GHz})/1\; {\rm \mu s})^{1/2} \nu_{\rm GHz}^{-2}$~cm where $\tau_{\rm sc}$ is the scattering time, $d_{\rm screen}$ the distance to the scattering screen, and $\nu_{\rm GHz}$ the frequency in gigahertz.  Some bursts have been constrained to have sub-microsecond gigahertz scattering times $\tau_{\rm sc}(1~{\rm GHz}) < 1\,\mu{\rm s}$, and a screen at $0.01\; {\rm pc}$ is motivated by calculations for the sizes of young magnetar nebulae \citep{Metzger:2019una}.

A second condition is that multipath propagation from strong scattering in our Galaxy does not decorrelate the fields. 
To match the $\sim 1 \, \nu_{\rm GHz}^{-4}\, \mu{\rm s}$ diffractive scattering times observed for sightlines orthogonal to the Milky Way disk, the effective radius of the scattered micro-images should be $R_{\rm sc} \sim 10^{12} \nu_{\rm GHz}^{-2}\,$cm. The zero-lag electric field correlations between the two unresolved lensed images will scale with the fraction by which their scattered images overlap or $\sim (R_{\rm sc}/[\theta_E\,d_{\rm MW}])^2 \sim 10^{-2} (G \mu/10^{-8})^{-2}\,\nu_{\rm GHz}^{-4}$ (as the two string images will essentially share diffractive paths where they overlap). The correlations are stronger at lower frequencies.  Thus, if $G \mu < 10^{-8}~\nu_{\rm GHz}^{-2}$, the the two lensed images can show appreciable electric field correlations. Alternatively, one could observe at $\nu_{\rm GHz} \gtrsim 5\,$GHz where Milky Way scattering becomes weak for sightlines out of the disk.

\subsection{Magnification Difference}
Ideally, the two images created by a straight cosmic string have exactly the same magnification factors. However, the large-scale matter inhomogeneities integrated along the line of sight result in small differential magnifications of the two lensed images through weak lensing effects.
The typical size of the magnification difference between the two images can be estimated using the angular correlation function of the weak lensing convergence, $\omega_{\kappa}(\theta)$. In the weak magnification regime, the magnification is determined from the convergence through $M = 1+2\,\kappa$.
Since we are interested in correlation on extremely small angular scales, the relevant matter power spectrum should be computed on a scale $\theta_E\,D\sim 100\,{\rm pc}$ for $G\mu \sim 10^{-8}$, where the characteristic line-of-sight distance $D$ is on the order of cosmological distances. 
This length scale is deep in the nonlinear regime beyond the reach of practical structure formation simulations, so for a crude estimate we choose to extrapolate the matter power spectrum using analytical models. We use the stable clustering model to compute the nonlinear power spectrum \cite{Smith_2003}.
The magnification difference due to the nonlinear structures therefore can be estimated as:
\begin{equation}
\begin{split}
        \Delta M &= 2\sqrt{|\omega_{\kappa}(2\theta_E)-\omega_{\kappa}(0)|},\\
        &\sim 2.4\times 10^{-3}\left(\frac{D}{{\rm Gpc}/h}\right)^{7/4}\left(\frac{\theta_E}{10^{-7}}\right)^{1/4}.
\end{split}
\end{equation}
We see that the magnification difference induced by weak lensing is negligibly small. 

Along the line of sight, stars that wander outside galaxies can cause weak lensing as well, with an average magnification difference approximately given by the energy fraction of stars outside galaxies in the Universe, $\Omega_{\star}\lesssim 10^{-4}$ ($\sim5\%$ of the matter is baryonic, $\sim5\%$ of the baryons are in stars, and less than $10\%$ of them are outside galaxies.). Because magnification from a point mass falls off as angle to the minus fourth power, lensing from stars can cause a magnification difference of $\gtrsim 10^{-2}$ with a probability of $10\,\Omega_{\star} \lesssim 0.1\%$. 
Therefore, the confounding effect of wandering stellar microlenses should be negligible, and we expect two equally bright images.

\subsection{Caustic lensing events}

A pair of equally bright lensed images can be naturally produced by ordinary astrophysical lenses (e.g. galaxies) if the source lies in the proximity of a lensing caustic and is hence highly magnified. The time delay between the highly magnified image pair can be orders of magnitude smaller than the Schwarzschild timescale corresponding to the lens mass, $\sim 10^7\,{\rm s}\,(M_L/10^{12}\,M_\odot)$ (we consider the typical galaxy lens mass scale, $M_L \sim 10^{12}\,M_\odot$, corresponding to an Einstein angle $\theta_E \sim 1''$). How do we distinguish cosmic string lensing from caustic lensing?

First, we should estimate how often caustic lensing occurs. For a lens with a smooth mass profile, this time delay is of order $\Delta t \sim D\,\theta^3/\theta_E$, where $\theta_E$ is the lens Einstein angular scale and $\theta$ is the angular separation of the image pair. To realize a given time delay $\Delta t$, the source position has to lie fortuitously within an angular distance $y$ to the lensing caustic, which is a fraction $y/\theta_E \sim (\theta/\theta_E)^2$ of the lens Einstein scale $\theta_E$. This means that
\begin{align}
    \frac{y}{\theta_E} \sim \left( \frac{\Delta t}{D\,\theta^2_E} \right)^{2/3} \approx 10^{-4}\,\left( \frac{\Delta t}{100\,{\rm s}} \right)^{2/3}\,\left( \frac{M_L}{10^{12}\,M_\odot} \right)^{-2/3}.
\end{align}
This requires fine-tuning of the source position, which must result in a high value for the magnification $\mu \sim (\theta/\theta_E)^{-1} \sim (y/\theta_E)^{-1/2}$. This can be rewritten as
\begin{align}
    \mu \sim 100\,\left( \frac{\Delta t}{100\,{\rm s}} \right)^{-1/3}\,\left( \frac{M_L}{10^{12}\,M_\odot} \right)^{1/3}.
\end{align}
Combining with Eq.(\ref{eq:timedelay}), we have
\begin{align}
\label{eq:mucaustic}
    \mu \sim 60\,\left(\frac{G\mu}{10^{-8}}\right)^{-2/3}\,\left(\frac{M_L}{10^{12}\,M_\odot} \right)^{1/3}\,\left( \frac{D}{{\rm Gpc}} \right)^{-1/3}.
\end{align}
FRBs lensed by galaxies must have magnifications as high as this to have time delays similar to that from string lensing.

Dominated by fold caustics, the cumulative probability for high magnification $\mu \gg 1$ should follow a universal power-law $P(>\mu) \propto 1/\mu^2$. Conservatively, we may consider FRB sources up to $z=5$ (see Fig.~(\ref{fig:pdf})) and have ~\cite{Dai:2016igl} (also see \cite{Hilbert:2007ny, Hilbert:2007jd, Oguri:2018muv})
\begin{align}
    P(>\mu) \simeq 8 \times 10^{-7}\,\left( \frac{\mu}{50} \right)^{-2},\quad {\rm for}\quad \mu > 50,
\end{align}
Inserting Eq.(\ref{eq:mucaustic}), we obtain the probability of confounding caustic lensing events
\begin{align}
\label{eq:Pmu}
    P(>\mu) \simeq 6 \times 10^{-7}\,\left(\frac{G\mu}{10^{-8}}\right)^{\frac43}\,\left(\frac{M_L}{10^{12}\,M_\odot} \right)^{-\frac23}\,\left( \frac{D}{{\rm Gpc}} \right)^{\frac23}.
\end{align}

Having a slightly steeper scaling with $G\mu$ than in Eq.(\ref{eq:detection_rate}), Eq.~(\ref{eq:Pmu}) falls below the string lensing probability $\gtrsim 2\times 10^{-6}\,(N_{\rm str}/30)\,(G\,\mu / 10^{-8})$ for $G\mu< 10^{-7}$. While a pair of equally bright FRB bursts separated by a few hundred seconds in time may still be a highly magnified caustic lensing event, we expect a significant excess in the caustic lensing rate, if string lenses do exist, at sufficiently high magnifications given by Eq.~(\ref{eq:mucaustic}).

We note that our simple estimate does not account for DM subhalos that reside within the lens galactic halo, which have significant perturbing effects in the proximity of a caustic~\cite{Dai:2018mxx, Dai:2020rio}. Such details are beyond the scope of this work, but should be investigated in future works.




If the lensed FRB source is localized to arcsecond scales, optical follow-up observations can be used to look for a galaxy-galaxy lensing system. The image separation for a caustic lensing event can be estimated as
\begin{align}
    \theta & \sim \left( \frac{\Delta t}{D\,\theta^2_E} \right)^{1/3}\,\theta_E, \nonumber\\
    & \sim 0.1''\,\left( \frac{\Delta t}{100\,{\rm s}} \right)^{1/3}\,\left( \frac{M_L}{10^{12}\,M_\odot} \right)^{1/6}\,\left( \frac{D}{{\rm Gpc}} \right)^{-1/2}.
\end{align}
If no caustic crossing galaxy is found at the location of the FRB source, the galaxy-galaxy lensing scenario can be ruled out.

Time delay variation is another potential observable of distinguishing power if the lensed FRB source repeats. For caustic lensing event, we estimate the time delay variation to be
\begin{align}
\label{eq:deltatcaustic}
    & \delta t \sim D\,\theta^2_E\,\left( \frac{y}{\theta_E\,D} \right)^{1/2}\,\left( \frac{v\,T_{\rm obs}}{\theta_E\,D} \right) \nonumber\\
    & \sim 0.04\,{\rm s}\,\left( \frac{\Delta t}{500\,{\rm s}} \right)^{1/3}\,\left( \frac{v}{10^3\,{\rm km\,s^{-1}}} \right)\,\left( \frac{T_{\rm obs}}{{\rm yr}} \right) \nonumber\\
    & \times \left( \frac{M_L}{10^{12}\,M_\odot} \right)^{1/6}\,\left( \frac{D}{{\rm Gpc}} \right)^{-1/2},
\end{align}
where $v$ is the typical relative transverse motion between the source and the galaxy lens, and $T_{\rm obs}$ is the observation time span. Cosmic strings move significantly faster than matter structures in the Universe such as galaxies. Through a comparison between Eq.(\ref{eq:deltatcaustic}) and Eq.(\ref{eq:deltat}), we see that a genuine string lensed FRB source should exhibit a larger time delay variation over the same observation time span, since cosmic strings move significantly faster than any galaxies in the Universe. 

\section{Conclusion}
\label{sec:concl}

We have shown that strong gravitational lensing of FRBs can probe cosmic string networks in the interesting parameter space in which the string tension is too small to produce discernible lensing signatures in the Cosmic Microwave Background anisotropies or in galaxy surveys. Two lensed images with a time delay on the order of a few hundred seconds are expected for string tension $G\mu\sim10^{-8}$. The two lensed images can be resolved in the time domain as the time delay is much longer than the FRB burst duration. Spatially resolving the two lensed images is not required, but can be achievable with VLBI observations. The time delay expected from string lensing is of a unique order of magnitude when compared with other strong lensing phenomena --- microlensing time delays due to stars or stellar remnants are much shorter, time delays of galaxy or galaxy cluster lensing are much longer. The magnification difference between the two lensed images due to line-of-sight weak lensing effects by the large scale structure is expected to be negligibly small --- an estimated fractional difference of only $\sim 10^{-3}$. This suggests that a lensed image pair are expected to show essentially the same magnification. A confounding image pair arising from the caustic proximity of a galaxy lens with a similar time delay is expected to be less probable.

We have computed the strong lensing rate as a function of the string tension, $G\mu$, and the number of FRBs detected in an FRB survey, $N_{\rm FRB}$. We have done so for much different FRB redshift distribution models in which the FRBs trace stellar mass and the star formation rate. The expected lensing rate differs by a factor of $\sim 2$ among these FRB population models. The stellar mass model predicts relatively low redshifts for the distribution of FRBs, which does not agree with CHIME if the detection efficiency is included.
We find an interesting strong lensing rate given the prospect of detecting a large number of FRBs ($N_{\rm FRB} \sim 10^6$--$10^8$) with future radio surveys --- the lensed event count reaches order unity if $G\mu\,N_{\rm FRB} \gtrsim 0.01$.
Assuming a 10-year observational span, our calculations suggest that future SKA surveys with up to $\sim 10^7$ FRB detections will enable us to probe string tensions as low as $G\,\mu \sim 10^{-9}$. This is close to the lowest tension value that can explain the candidate stochastic gravitational wave background uncovered by pulsar timing arrays \cite{Blasi:2020mfx}. Therefore, future FRB surveys have the potential to discover relic cosmic strings from the early Universe.

We have pointed out that it is possible to robustly establish that two signals are genuine lensed images that come from the same FRB burst. This is because each radio burst has a unique fingerprint spectral energy distribution. Furthermore, it may be feasible to correlate the electric field time series to judge if the two lensed images originate from the same source.

If the string lenses are associated with an axion-like particle that couples to electromagnetism, they are expected to cause cosmic birefringence in which the polarization states of the CMB photons are rotated. A unique discontinuity in the spatial pattern of the polarization rotation angle is expected across the string location in the sky \cite{Agrawal:2019lkr, Jain:2021shf}. Detecting just one string lensed FRB event will pinpoint the sky location of a string segment, which will facilitate targeted search of nearby string signatures, either through birefringence in the CMB \cite{Yin:2021kmx}, or through multiply-imaged lensing of other background sources along the string.\\

\section*{Acknowledgements}
HX is supported in part by the U.S. Department of Energy under grant number DE-SC0011637. LD acknowledges the research grant support from the Alfred P.
Sloan Foundation (award number FG-2021-16495). We acknowledge support from NSF award AST-2007012
\bibliographystyle{apsrev4-2}
\bibliography{cosmic_string}

\begin{thebibliography}{76}%
\makeatletter
\providecommand \@ifxundefined [1]{%
 \@ifx{#1\undefined}
}%
\providecommand \@ifnum [1]{%
 \ifnum #1\expandafter \@firstoftwo
 \else \expandafter \@secondoftwo
 \fi
}%
\providecommand \@ifx [1]{%
 \ifx #1\expandafter \@firstoftwo
 \else \expandafter \@secondoftwo
 \fi
}%
\providecommand \natexlab [1]{#1}%
\providecommand \enquote  [1]{``#1''}%
\providecommand \bibnamefont  [1]{#1}%
\providecommand \bibfnamefont [1]{#1}%
\providecommand \citenamefont [1]{#1}%
\providecommand \href@noop [0]{\@secondoftwo}%
\providecommand \href [0]{\begingroup \@sanitize@url \@href}%
\providecommand \@href[1]{\@@startlink{#1}\@@href}%
\providecommand \@@href[1]{\endgroup#1\@@endlink}%
\providecommand \@sanitize@url [0]{\catcode `\\12\catcode `\$12\catcode
  `\&12\catcode `\#12\catcode `\^12\catcode `\_12\catcode `\%12\relax}%
\providecommand \@@startlink[1]{}%
\providecommand \@@endlink[0]{}%
\providecommand \url  [0]{\begingroup\@sanitize@url \@url }%
\providecommand \@url [1]{\endgroup\@href {#1}{\urlprefix }}%
\providecommand \urlprefix  [0]{URL }%
\providecommand \Eprint [0]{\href }%
\providecommand \doibase [0]{https://doi.org/}%
\providecommand \selectlanguage [0]{\@gobble}%
\providecommand \bibinfo  [0]{\@secondoftwo}%
\providecommand \bibfield  [0]{\@secondoftwo}%
\providecommand \translation [1]{[#1]}%
\providecommand \BibitemOpen [0]{}%
\providecommand \bibitemStop [0]{}%
\providecommand \bibitemNoStop [0]{.\EOS\space}%
\providecommand \EOS [0]{\spacefactor3000\relax}%
\providecommand \BibitemShut  [1]{\csname bibitem#1\endcsname}%
\let\auto@bib@innerbib\@empty
\bibitem [{\citenamefont {Kibble}(1976)}]{Kibble:1976sj}%
  \BibitemOpen
  \bibfield  {author} {\bibinfo {author} {\bibfnamefont {T.~W.~B.}\
  \bibnamefont {Kibble}},\ }\href {https://doi.org/10.1088/0305-4470/9/8/029}
  {\bibfield  {journal} {\bibinfo  {journal} {J. Phys. A}\ }\textbf {\bibinfo
  {volume} {9}},\ \bibinfo {pages} {1387} (\bibinfo {year} {1976})}\BibitemShut
  {NoStop}%
\bibitem [{\citenamefont {Hindmarsh}\ and\ \citenamefont
  {Kibble}(1995)}]{Hindmarsh:1994re}%
  \BibitemOpen
  \bibfield  {author} {\bibinfo {author} {\bibfnamefont {M.~B.}\ \bibnamefont
  {Hindmarsh}}\ and\ \bibinfo {author} {\bibfnamefont {T.~W.~B.}\ \bibnamefont
  {Kibble}},\ }\href {https://doi.org/10.1088/0034-4885/58/5/001} {\bibfield
  {journal} {\bibinfo  {journal} {Rept. Prog. Phys.}\ }\textbf {\bibinfo
  {volume} {58}},\ \bibinfo {pages} {477} (\bibinfo {year} {1995})},\ \Eprint
  {https://arxiv.org/abs/hep-ph/9411342} {arXiv:hep-ph/9411342} \BibitemShut
  {NoStop}%
\bibitem [{\citenamefont {Vilenkin}(1985)}]{Vilenkin:1984ib}%
  \BibitemOpen
  \bibfield  {author} {\bibinfo {author} {\bibfnamefont {A.}~\bibnamefont
  {Vilenkin}},\ }\href {https://doi.org/10.1016/0370-1573(85)90033-X}
  {\bibfield  {journal} {\bibinfo  {journal} {Phys. Rept.}\ }\textbf {\bibinfo
  {volume} {121}},\ \bibinfo {pages} {263} (\bibinfo {year}
  {1985})}\BibitemShut {NoStop}%
\bibitem [{\citenamefont {Dvali}\ and\ \citenamefont
  {Tye}(1999)}]{Dvali:1998pa}%
  \BibitemOpen
  \bibfield  {author} {\bibinfo {author} {\bibfnamefont {G.~R.}\ \bibnamefont
  {Dvali}}\ and\ \bibinfo {author} {\bibfnamefont {S.~H.~H.}\ \bibnamefont
  {Tye}},\ }\href {https://doi.org/10.1016/S0370-2693(99)00132-X} {\bibfield
  {journal} {\bibinfo  {journal} {Phys. Lett. B}\ }\textbf {\bibinfo {volume}
  {450}},\ \bibinfo {pages} {72} (\bibinfo {year} {1999})},\ \Eprint
  {https://arxiv.org/abs/hep-ph/9812483} {arXiv:hep-ph/9812483} \BibitemShut
  {NoStop}%
\bibitem [{\citenamefont {Baumann}\ \emph {et~al.}(2008)\citenamefont
  {Baumann}, \citenamefont {Dymarsky}, \citenamefont {Klebanov},\ and\
  \citenamefont {McAllister}}]{Baumann:2007ah}%
  \BibitemOpen
  \bibfield  {author} {\bibinfo {author} {\bibfnamefont {D.}~\bibnamefont
  {Baumann}}, \bibinfo {author} {\bibfnamefont {A.}~\bibnamefont {Dymarsky}},
  \bibinfo {author} {\bibfnamefont {I.~R.}\ \bibnamefont {Klebanov}},\ and\
  \bibinfo {author} {\bibfnamefont {L.}~\bibnamefont {McAllister}},\ }\href
  {https://doi.org/10.1088/1475-7516/2008/01/024} {\bibfield  {journal}
  {\bibinfo  {journal} {JCAP}\ }\textbf {\bibinfo {volume} {01}},\ \bibinfo
  {pages} {024}},\ \Eprint {https://arxiv.org/abs/0706.0360} {arXiv:0706.0360
  [hep-th]} \BibitemShut {NoStop}%
\bibitem [{\citenamefont {Burgess}\ \emph {et~al.}(2001)\citenamefont
  {Burgess}, \citenamefont {Majumdar}, \citenamefont {Nolte}, \citenamefont
  {Quevedo}, \citenamefont {Rajesh},\ and\ \citenamefont
  {Zhang}}]{Burgess:2001fx}%
  \BibitemOpen
  \bibfield  {author} {\bibinfo {author} {\bibfnamefont {C.~P.}\ \bibnamefont
  {Burgess}}, \bibinfo {author} {\bibfnamefont {M.}~\bibnamefont {Majumdar}},
  \bibinfo {author} {\bibfnamefont {D.}~\bibnamefont {Nolte}}, \bibinfo
  {author} {\bibfnamefont {F.}~\bibnamefont {Quevedo}}, \bibinfo {author}
  {\bibfnamefont {G.}~\bibnamefont {Rajesh}},\ and\ \bibinfo {author}
  {\bibfnamefont {R.-J.}\ \bibnamefont {Zhang}},\ }\href
  {https://doi.org/10.1088/1126-6708/2001/07/047} {\bibfield  {journal}
  {\bibinfo  {journal} {JHEP}\ }\textbf {\bibinfo {volume} {07}},\ \bibinfo
  {pages} {047}},\ \Eprint {https://arxiv.org/abs/hep-th/0105204}
  {arXiv:hep-th/0105204} \BibitemShut {NoStop}%
\bibitem [{\citenamefont {Kibble}\ \emph {et~al.}(1982)\citenamefont {Kibble},
  \citenamefont {Lazarides},\ and\ \citenamefont {Shafi}}]{Kibble:1982ae}%
  \BibitemOpen
  \bibfield  {author} {\bibinfo {author} {\bibfnamefont {T.~W.~B.}\
  \bibnamefont {Kibble}}, \bibinfo {author} {\bibfnamefont {G.}~\bibnamefont
  {Lazarides}},\ and\ \bibinfo {author} {\bibfnamefont {Q.}~\bibnamefont
  {Shafi}},\ }\href {https://doi.org/10.1016/0370-2693(82)90829-2} {\bibfield
  {journal} {\bibinfo  {journal} {Phys. Lett. B}\ }\textbf {\bibinfo {volume}
  {113}},\ \bibinfo {pages} {237} (\bibinfo {year} {1982})}\BibitemShut
  {NoStop}%
\bibitem [{\citenamefont {Jeannerot}\ \emph {et~al.}(2003)\citenamefont
  {Jeannerot}, \citenamefont {Rocher},\ and\ \citenamefont
  {Sakellariadou}}]{Jeannerot:2003qv}%
  \BibitemOpen
  \bibfield  {author} {\bibinfo {author} {\bibfnamefont {R.}~\bibnamefont
  {Jeannerot}}, \bibinfo {author} {\bibfnamefont {J.}~\bibnamefont {Rocher}},\
  and\ \bibinfo {author} {\bibfnamefont {M.}~\bibnamefont {Sakellariadou}},\
  }\href {https://doi.org/10.1103/PhysRevD.68.103514} {\bibfield  {journal}
  {\bibinfo  {journal} {Phys. Rev. D}\ }\textbf {\bibinfo {volume} {68}},\
  \bibinfo {pages} {103514} (\bibinfo {year} {2003})},\ \Eprint
  {https://arxiv.org/abs/hep-ph/0308134} {arXiv:hep-ph/0308134} \BibitemShut
  {NoStop}%
\bibitem [{\citenamefont {Vilenkin}\ and\ \citenamefont
  {Everett}(1982)}]{Vilenkin:1982ks}%
  \BibitemOpen
  \bibfield  {author} {\bibinfo {author} {\bibfnamefont {A.}~\bibnamefont
  {Vilenkin}}\ and\ \bibinfo {author} {\bibfnamefont {A.~E.}\ \bibnamefont
  {Everett}},\ }\href {https://doi.org/10.1103/PhysRevLett.48.1867} {\bibfield
  {journal} {\bibinfo  {journal} {Phys. Rev. Lett.}\ }\textbf {\bibinfo
  {volume} {48}},\ \bibinfo {pages} {1867} (\bibinfo {year}
  {1982})}\BibitemShut {NoStop}%
\bibitem [{\citenamefont {Blanco-Pillado}\ \emph
  {et~al.}(2011{\natexlab{a}})\citenamefont {Blanco-Pillado}, \citenamefont
  {Olum},\ and\ \citenamefont {Shlaer}}]{PhysRevD.83.083514}%
  \BibitemOpen
  \bibfield  {author} {\bibinfo {author} {\bibfnamefont {J.~J.}\ \bibnamefont
  {Blanco-Pillado}}, \bibinfo {author} {\bibfnamefont {K.~D.}\ \bibnamefont
  {Olum}},\ and\ \bibinfo {author} {\bibfnamefont {B.}~\bibnamefont {Shlaer}},\
  }\href {https://doi.org/10.1103/PhysRevD.83.083514} {\bibfield  {journal}
  {\bibinfo  {journal} {Phys. Rev. D}\ }\textbf {\bibinfo {volume} {83}},\
  \bibinfo {pages} {083514} (\bibinfo {year} {2011}{\natexlab{a}})}\BibitemShut
  {NoStop}%
\bibitem [{\citenamefont {Arzoumanian}\ \emph {et~al.}(2020)\citenamefont
  {Arzoumanian} \emph {et~al.}}]{NANOGrav:2020bcs}%
  \BibitemOpen
  \bibfield  {author} {\bibinfo {author} {\bibfnamefont {Z.}~\bibnamefont
  {Arzoumanian}} \emph {et~al.} (\bibinfo {collaboration} {NANOGrav}),\ }\href
  {https://doi.org/10.3847/2041-8213/abd401} {\bibfield  {journal} {\bibinfo
  {journal} {Astrophys. J. Lett.}\ }\textbf {\bibinfo {volume} {905}},\
  \bibinfo {pages} {L34} (\bibinfo {year} {2020})},\ \Eprint
  {https://arxiv.org/abs/2009.04496} {arXiv:2009.04496 [astro-ph.HE]}
  \BibitemShut {NoStop}%
\bibitem [{\citenamefont {Blasi}\ \emph {et~al.}(2021)\citenamefont {Blasi},
  \citenamefont {Brdar},\ and\ \citenamefont {Schmitz}}]{Blasi:2020mfx}%
  \BibitemOpen
  \bibfield  {author} {\bibinfo {author} {\bibfnamefont {S.}~\bibnamefont
  {Blasi}}, \bibinfo {author} {\bibfnamefont {V.}~\bibnamefont {Brdar}},\ and\
  \bibinfo {author} {\bibfnamefont {K.}~\bibnamefont {Schmitz}},\ }\href
  {https://doi.org/10.1103/PhysRevLett.126.041305} {\bibfield  {journal}
  {\bibinfo  {journal} {Phys. Rev. Lett.}\ }\textbf {\bibinfo {volume} {126}},\
  \bibinfo {pages} {041305} (\bibinfo {year} {2021})},\ \Eprint
  {https://arxiv.org/abs/2009.06607} {arXiv:2009.06607 [astro-ph.CO]}
  \BibitemShut {NoStop}%
\bibitem [{\citenamefont {Ellis}\ and\ \citenamefont
  {Lewicki}(2021)}]{Ellis:2020ena}%
  \BibitemOpen
  \bibfield  {author} {\bibinfo {author} {\bibfnamefont {J.}~\bibnamefont
  {Ellis}}\ and\ \bibinfo {author} {\bibfnamefont {M.}~\bibnamefont
  {Lewicki}},\ }\href {https://doi.org/10.1103/PhysRevLett.126.041304}
  {\bibfield  {journal} {\bibinfo  {journal} {Phys. Rev. Lett.}\ }\textbf
  {\bibinfo {volume} {126}},\ \bibinfo {pages} {041304} (\bibinfo {year}
  {2021})},\ \Eprint {https://arxiv.org/abs/2009.06555} {arXiv:2009.06555
  [astro-ph.CO]} \BibitemShut {NoStop}%
\bibitem [{\citenamefont {Buchmuller}\ \emph {et~al.}(2020)\citenamefont
  {Buchmuller}, \citenamefont {Domcke},\ and\ \citenamefont
  {Schmitz}}]{Buchmuller:2020lbh}%
  \BibitemOpen
  \bibfield  {author} {\bibinfo {author} {\bibfnamefont {W.}~\bibnamefont
  {Buchmuller}}, \bibinfo {author} {\bibfnamefont {V.}~\bibnamefont {Domcke}},\
  and\ \bibinfo {author} {\bibfnamefont {K.}~\bibnamefont {Schmitz}},\ }\href
  {https://doi.org/10.1016/j.physletb.2020.135914} {\bibfield  {journal}
  {\bibinfo  {journal} {Phys. Lett. B}\ }\textbf {\bibinfo {volume} {811}},\
  \bibinfo {pages} {135914} (\bibinfo {year} {2020})},\ \Eprint
  {https://arxiv.org/abs/2009.10649} {arXiv:2009.10649 [astro-ph.CO]}
  \BibitemShut {NoStop}%
\bibitem [{\citenamefont {Bogomolny}(1976)}]{Bogomolny:1975de}%
  \BibitemOpen
  \bibfield  {author} {\bibinfo {author} {\bibfnamefont {E.~B.}\ \bibnamefont
  {Bogomolny}},\ }\href@noop {} {\bibfield  {journal} {\bibinfo  {journal}
  {Sov. J. Nucl. Phys.}\ }\textbf {\bibinfo {volume} {24}},\ \bibinfo {pages}
  {449} (\bibinfo {year} {1976})}\BibitemShut {NoStop}%
\bibitem [{\citenamefont {King}\ \emph {et~al.}(2021)\citenamefont {King},
  \citenamefont {Pascoli}, \citenamefont {Turner},\ and\ \citenamefont
  {Zhou}}]{King:2020hyd}%
  \BibitemOpen
  \bibfield  {author} {\bibinfo {author} {\bibfnamefont {S.~F.}\ \bibnamefont
  {King}}, \bibinfo {author} {\bibfnamefont {S.}~\bibnamefont {Pascoli}},
  \bibinfo {author} {\bibfnamefont {J.}~\bibnamefont {Turner}},\ and\ \bibinfo
  {author} {\bibfnamefont {Y.-L.}\ \bibnamefont {Zhou}},\ }\href
  {https://doi.org/10.1103/PhysRevLett.126.021802} {\bibfield  {journal}
  {\bibinfo  {journal} {Phys. Rev. Lett.}\ }\textbf {\bibinfo {volume} {126}},\
  \bibinfo {pages} {021802} (\bibinfo {year} {2021})},\ \Eprint
  {https://arxiv.org/abs/2005.13549} {arXiv:2005.13549 [hep-ph]} \BibitemShut
  {NoStop}%
\bibitem [{\citenamefont {Lyth}\ and\ \citenamefont
  {Riotto}(1999)}]{Lyth:1998xn}%
  \BibitemOpen
  \bibfield  {author} {\bibinfo {author} {\bibfnamefont {D.~H.}\ \bibnamefont
  {Lyth}}\ and\ \bibinfo {author} {\bibfnamefont {A.}~\bibnamefont {Riotto}},\
  }\href {https://doi.org/10.1016/S0370-1573(98)00128-8} {\bibfield  {journal}
  {\bibinfo  {journal} {Phys. Rept.}\ }\textbf {\bibinfo {volume} {314}},\
  \bibinfo {pages} {1} (\bibinfo {year} {1999})},\ \Eprint
  {https://arxiv.org/abs/hep-ph/9807278} {arXiv:hep-ph/9807278} \BibitemShut
  {NoStop}%
\bibitem [{\citenamefont {Turok}(1984)}]{Turok:1984cn}%
  \BibitemOpen
  \bibfield  {author} {\bibinfo {author} {\bibfnamefont {N.}~\bibnamefont
  {Turok}},\ }\href {https://doi.org/10.1016/0550-3213(84)90407-3} {\bibfield
  {journal} {\bibinfo  {journal} {Nucl. Phys. B}\ }\textbf {\bibinfo {volume}
  {242}},\ \bibinfo {pages} {520} (\bibinfo {year} {1984})}\BibitemShut
  {NoStop}%
\bibitem [{\citenamefont {Vilenkin}\ and\ \citenamefont
  {Shafi}(1983)}]{Vilenkin:1983jv}%
  \BibitemOpen
  \bibfield  {author} {\bibinfo {author} {\bibfnamefont {A.}~\bibnamefont
  {Vilenkin}}\ and\ \bibinfo {author} {\bibfnamefont {Q.}~\bibnamefont
  {Shafi}},\ }\href {https://doi.org/10.1103/PhysRevLett.51.1716} {\bibfield
  {journal} {\bibinfo  {journal} {Phys. Rev. Lett.}\ }\textbf {\bibinfo
  {volume} {51}},\ \bibinfo {pages} {1716} (\bibinfo {year}
  {1983})}\BibitemShut {NoStop}%
\bibitem [{\citenamefont {{Liddle}}\ and\ \citenamefont
  {{Lyth}}(2000)}]{2000cils.book.....L}%
  \BibitemOpen
  \bibfield  {author} {\bibinfo {author} {\bibfnamefont {A.~R.}\ \bibnamefont
  {{Liddle}}}\ and\ \bibinfo {author} {\bibfnamefont {D.~H.}\ \bibnamefont
  {{Lyth}}},\ }\href@noop {} {\emph {\bibinfo {title} {{Cosmological Inflation
  and Large-Scale Structure}}}}\ (\bibinfo {year} {2000})\BibitemShut {NoStop}%
\bibitem [{\citenamefont {Kaiser}\ and\ \citenamefont
  {Stebbins}(1984)}]{Kaiser:1984iv}%
  \BibitemOpen
  \bibfield  {author} {\bibinfo {author} {\bibfnamefont {N.}~\bibnamefont
  {Kaiser}}\ and\ \bibinfo {author} {\bibfnamefont {A.}~\bibnamefont
  {Stebbins}},\ }\href {https://doi.org/10.1038/310391a0} {\bibfield  {journal}
  {\bibinfo  {journal} {Nature}\ }\textbf {\bibinfo {volume} {310}},\ \bibinfo
  {pages} {391} (\bibinfo {year} {1984})}\BibitemShut {NoStop}%
\bibitem [{\citenamefont {Charnock}\ \emph {et~al.}(2016)\citenamefont
  {Charnock}, \citenamefont {Avgoustidis}, \citenamefont {Copeland},\ and\
  \citenamefont {Moss}}]{Charnock:2016nzm}%
  \BibitemOpen
  \bibfield  {author} {\bibinfo {author} {\bibfnamefont {T.}~\bibnamefont
  {Charnock}}, \bibinfo {author} {\bibfnamefont {A.}~\bibnamefont
  {Avgoustidis}}, \bibinfo {author} {\bibfnamefont {E.~J.}\ \bibnamefont
  {Copeland}},\ and\ \bibinfo {author} {\bibfnamefont {A.}~\bibnamefont
  {Moss}},\ }\href {https://doi.org/10.1103/PhysRevD.93.123503} {\bibfield
  {journal} {\bibinfo  {journal} {Phys. Rev. D}\ }\textbf {\bibinfo {volume}
  {93}},\ \bibinfo {pages} {123503} (\bibinfo {year} {2016})},\ \Eprint
  {https://arxiv.org/abs/1603.01275} {arXiv:1603.01275 [astro-ph.CO]}
  \BibitemShut {NoStop}%
\bibitem [{\citenamefont {Ade}\ \emph {et~al.}(2016)\citenamefont {Ade} \emph
  {et~al.}}]{Planck:2015fie}%
  \BibitemOpen
  \bibfield  {author} {\bibinfo {author} {\bibfnamefont {P.~A.~R.}\
  \bibnamefont {Ade}} \emph {et~al.} (\bibinfo {collaboration} {Planck}),\
  }\href {https://doi.org/10.1051/0004-6361/201525830} {\bibfield  {journal}
  {\bibinfo  {journal} {Astron. Astrophys.}\ }\textbf {\bibinfo {volume}
  {594}},\ \bibinfo {pages} {A13} (\bibinfo {year} {2016})},\ \Eprint
  {https://arxiv.org/abs/1502.01589} {arXiv:1502.01589 [astro-ph.CO]}
  \BibitemShut {NoStop}%
\bibitem [{\citenamefont {Sazhin}\ \emph {et~al.}(2007)\citenamefont {Sazhin},
  \citenamefont {Khovanskaya}, \citenamefont {Capaccioli}, \citenamefont
  {Longo}, \citenamefont {Paolillo}, \citenamefont {Covone}, \citenamefont
  {Grogin},\ and\ \citenamefont {Schreier}}]{Sazhin:2006kf}%
  \BibitemOpen
  \bibfield  {author} {\bibinfo {author} {\bibfnamefont {M.~V.}\ \bibnamefont
  {Sazhin}}, \bibinfo {author} {\bibfnamefont {O.~S.}\ \bibnamefont
  {Khovanskaya}}, \bibinfo {author} {\bibfnamefont {M.}~\bibnamefont
  {Capaccioli}}, \bibinfo {author} {\bibfnamefont {G.}~\bibnamefont {Longo}},
  \bibinfo {author} {\bibfnamefont {M.}~\bibnamefont {Paolillo}}, \bibinfo
  {author} {\bibfnamefont {G.}~\bibnamefont {Covone}}, \bibinfo {author}
  {\bibfnamefont {N.~A.}\ \bibnamefont {Grogin}},\ and\ \bibinfo {author}
  {\bibfnamefont {E.~J.}\ \bibnamefont {Schreier}},\ }\href
  {https://doi.org/10.1111/j.1365-2966.2007.11543.x} {\bibfield  {journal}
  {\bibinfo  {journal} {Mon. Not. Roy. Astron. Soc.}\ }\textbf {\bibinfo
  {volume} {376}},\ \bibinfo {pages} {1731} (\bibinfo {year} {2007})},\ \Eprint
  {https://arxiv.org/abs/astro-ph/0611744} {arXiv:astro-ph/0611744}
  \BibitemShut {NoStop}%
\bibitem [{\citenamefont {Morganson}\ \emph {et~al.}(2010)\citenamefont
  {Morganson}, \citenamefont {Marshall}, \citenamefont {Treu}, \citenamefont
  {Schrabback},\ and\ \citenamefont {Blandford}}]{Morganson_2010}%
  \BibitemOpen
  \bibfield  {author} {\bibinfo {author} {\bibfnamefont {E.}~\bibnamefont
  {Morganson}}, \bibinfo {author} {\bibfnamefont {P.}~\bibnamefont {Marshall}},
  \bibinfo {author} {\bibfnamefont {T.}~\bibnamefont {Treu}}, \bibinfo {author}
  {\bibfnamefont {T.}~\bibnamefont {Schrabback}},\ and\ \bibinfo {author}
  {\bibfnamefont {R.~D.}\ \bibnamefont {Blandford}},\ }\href
  {https://doi.org/10.1111/j.1365-2966.2010.16562.x} {\bibfield  {journal}
  {\bibinfo  {journal} {Monthly Notices of the Royal Astronomical Society}\
  }\textbf {\bibinfo {volume} {406}},\ \bibinfo {pages} {2452} (\bibinfo {year}
  {2010})}\BibitemShut {NoStop}%
\bibitem [{\citenamefont {Christiansen}\ \emph {et~al.}(2011)\citenamefont
  {Christiansen}, \citenamefont {Albin}, \citenamefont {Fletcher},
  \citenamefont {Goldman}, \citenamefont {Teng}, \citenamefont {Foley},\ and\
  \citenamefont {Smoot}}]{Christiansen_2011}%
  \BibitemOpen
  \bibfield  {author} {\bibinfo {author} {\bibfnamefont {J.~L.}\ \bibnamefont
  {Christiansen}}, \bibinfo {author} {\bibfnamefont {E.}~\bibnamefont {Albin}},
  \bibinfo {author} {\bibfnamefont {T.}~\bibnamefont {Fletcher}}, \bibinfo
  {author} {\bibfnamefont {J.}~\bibnamefont {Goldman}}, \bibinfo {author}
  {\bibfnamefont {I.~P.~W.}\ \bibnamefont {Teng}}, \bibinfo {author}
  {\bibfnamefont {M.}~\bibnamefont {Foley}},\ and\ \bibinfo {author}
  {\bibfnamefont {G.~F.}\ \bibnamefont {Smoot}},\ }\bibfield  {journal}
  {\bibinfo  {journal} {Physical Review D}\ }\textbf {\bibinfo {volume} {83}},\
  \href {https://doi.org/10.1103/physrevd.83.122004}
  {10.1103/physrevd.83.122004} (\bibinfo {year} {2011})\BibitemShut {NoStop}%
\bibitem [{\citenamefont {Gasparini}\ \emph {et~al.}(2008)\citenamefont
  {Gasparini}, \citenamefont {Marshall}, \citenamefont {Treu}, \citenamefont
  {Morganson},\ and\ \citenamefont {Dubath}}]{Gasparini:2007jj}%
  \BibitemOpen
  \bibfield  {author} {\bibinfo {author} {\bibfnamefont {M.~A.}\ \bibnamefont
  {Gasparini}}, \bibinfo {author} {\bibfnamefont {P.}~\bibnamefont {Marshall}},
  \bibinfo {author} {\bibfnamefont {T.}~\bibnamefont {Treu}}, \bibinfo {author}
  {\bibfnamefont {E.}~\bibnamefont {Morganson}},\ and\ \bibinfo {author}
  {\bibfnamefont {F.}~\bibnamefont {Dubath}},\ }\href
  {https://doi.org/10.1111/j.1365-2966.2007.12657.x} {\bibfield  {journal}
  {\bibinfo  {journal} {Mon. Not. Roy. Astron. Soc.}\ }\textbf {\bibinfo
  {volume} {385}},\ \bibinfo {pages} {1959} (\bibinfo {year} {2008})},\ \Eprint
  {https://arxiv.org/abs/0710.5544} {arXiv:0710.5544 [astro-ph]} \BibitemShut
  {NoStop}%
\bibitem [{\citenamefont {{Vanderlinde}}\ \emph {et~al.}(2019)\citenamefont
  {{Vanderlinde}}, \citenamefont {{Liu}}, \citenamefont {{Gaensler}},
  \citenamefont {{Bond}}, \citenamefont {{Hinshaw}}, \citenamefont {{Ng}},
  \citenamefont {{Chiang}}, \citenamefont {{Stairs}}, \citenamefont {{Brown}},
  \citenamefont {{Sievers}}, \citenamefont {{Mena}}, \citenamefont {{Smith}},
  \citenamefont {{Bandura}}, \citenamefont {{Masui}}, \citenamefont
  {{Spekkens}}, \citenamefont {{Belostotski}}, \citenamefont {{Dobbs}},
  \citenamefont {{Turok}}, \citenamefont {{Boyle}}, \citenamefont {{Rupen}},
  \citenamefont {{Landecker}}, \citenamefont {{Pen}},\ and\ \citenamefont
  {{Kaspi}}}]{2019clrp.2020...28V}%
  \BibitemOpen
  \bibfield  {author} {\bibinfo {author} {\bibfnamefont {K.}~\bibnamefont
  {{Vanderlinde}}}, \bibinfo {author} {\bibfnamefont {A.}~\bibnamefont
  {{Liu}}}, \bibinfo {author} {\bibfnamefont {B.}~\bibnamefont {{Gaensler}}},
  \bibinfo {author} {\bibfnamefont {D.}~\bibnamefont {{Bond}}}, \bibinfo
  {author} {\bibfnamefont {G.}~\bibnamefont {{Hinshaw}}}, \bibinfo {author}
  {\bibfnamefont {C.}~\bibnamefont {{Ng}}}, \bibinfo {author} {\bibfnamefont
  {C.}~\bibnamefont {{Chiang}}}, \bibinfo {author} {\bibfnamefont
  {I.}~\bibnamefont {{Stairs}}}, \bibinfo {author} {\bibfnamefont {J.-A.}\
  \bibnamefont {{Brown}}}, \bibinfo {author} {\bibfnamefont {J.}~\bibnamefont
  {{Sievers}}}, \bibinfo {author} {\bibfnamefont {J.}~\bibnamefont {{Mena}}},
  \bibinfo {author} {\bibfnamefont {K.}~\bibnamefont {{Smith}}}, \bibinfo
  {author} {\bibfnamefont {K.}~\bibnamefont {{Bandura}}}, \bibinfo {author}
  {\bibfnamefont {K.}~\bibnamefont {{Masui}}}, \bibinfo {author} {\bibfnamefont
  {K.}~\bibnamefont {{Spekkens}}}, \bibinfo {author} {\bibfnamefont
  {L.}~\bibnamefont {{Belostotski}}}, \bibinfo {author} {\bibfnamefont
  {M.}~\bibnamefont {{Dobbs}}}, \bibinfo {author} {\bibfnamefont
  {N.}~\bibnamefont {{Turok}}}, \bibinfo {author} {\bibfnamefont
  {P.}~\bibnamefont {{Boyle}}}, \bibinfo {author} {\bibfnamefont
  {M.}~\bibnamefont {{Rupen}}}, \bibinfo {author} {\bibfnamefont
  {T.}~\bibnamefont {{Landecker}}}, \bibinfo {author} {\bibfnamefont {U.-L.}\
  \bibnamefont {{Pen}}},\ and\ \bibinfo {author} {\bibfnamefont
  {V.}~\bibnamefont {{Kaspi}}},\ }in\ \href
  {https://doi.org/10.5281/zenodo.3765414} {\emph {\bibinfo {booktitle}
  {Canadian Long Range Plan for Astronomy and Astrophysics White Papers}}},\
  Vol.\ \bibinfo {volume} {2020}\ (\bibinfo {year} {2019})\ p.~\bibinfo {pages}
  {28},\ \Eprint {https://arxiv.org/abs/1911.01777} {arXiv:1911.01777
  [astro-ph.IM]} \BibitemShut {NoStop}%
\bibitem [{\citenamefont {{Hallinan}}\ \emph {et~al.}(2019)\citenamefont
  {{Hallinan}}, \citenamefont {{Ravi}}, \citenamefont {{Weinreb}},
  \citenamefont {{Kocz}}, \citenamefont {{Huang}}, \citenamefont {{Woody}},
  \citenamefont {{Lamb}}, \citenamefont {{D'Addario}}, \citenamefont {{Catha}},
  \citenamefont {{Law}}, \citenamefont {{Kulkarni}}, \citenamefont {{Phinney}},
  \citenamefont {{Eastwood}}, \citenamefont {{Bouman}}, \citenamefont
  {{McLaughlin}}, \citenamefont {{Ransom}}, \citenamefont {{Siemens}},
  \citenamefont {{Cordes}}, \citenamefont {{Lynch}}, \citenamefont {{Kaplan}},
  \citenamefont {{Brazier}}, \citenamefont {{Bhatnagar}}, \citenamefont
  {{Myers}}, \citenamefont {{Walter}},\ and\ \citenamefont
  {{Gaensler}}}]{2019BAAS...51g.255H}%
  \BibitemOpen
  \bibfield  {author} {\bibinfo {author} {\bibfnamefont {G.}~\bibnamefont
  {{Hallinan}}}, \bibinfo {author} {\bibfnamefont {V.}~\bibnamefont {{Ravi}}},
  \bibinfo {author} {\bibfnamefont {S.}~\bibnamefont {{Weinreb}}}, \bibinfo
  {author} {\bibfnamefont {J.}~\bibnamefont {{Kocz}}}, \bibinfo {author}
  {\bibfnamefont {Y.}~\bibnamefont {{Huang}}}, \bibinfo {author} {\bibfnamefont
  {D.~P.}\ \bibnamefont {{Woody}}}, \bibinfo {author} {\bibfnamefont
  {J.}~\bibnamefont {{Lamb}}}, \bibinfo {author} {\bibfnamefont
  {L.}~\bibnamefont {{D'Addario}}}, \bibinfo {author} {\bibfnamefont
  {M.}~\bibnamefont {{Catha}}}, \bibinfo {author} {\bibfnamefont
  {C.}~\bibnamefont {{Law}}}, \bibinfo {author} {\bibfnamefont {S.~R.}\
  \bibnamefont {{Kulkarni}}}, \bibinfo {author} {\bibfnamefont {E.~S.}\
  \bibnamefont {{Phinney}}}, \bibinfo {author} {\bibfnamefont {M.~W.}\
  \bibnamefont {{Eastwood}}}, \bibinfo {author} {\bibfnamefont
  {K.}~\bibnamefont {{Bouman}}}, \bibinfo {author} {\bibfnamefont
  {M.}~\bibnamefont {{McLaughlin}}}, \bibinfo {author} {\bibfnamefont
  {S.}~\bibnamefont {{Ransom}}}, \bibinfo {author} {\bibfnamefont
  {X.}~\bibnamefont {{Siemens}}}, \bibinfo {author} {\bibfnamefont
  {J.}~\bibnamefont {{Cordes}}}, \bibinfo {author} {\bibfnamefont
  {R.}~\bibnamefont {{Lynch}}}, \bibinfo {author} {\bibfnamefont
  {D.}~\bibnamefont {{Kaplan}}}, \bibinfo {author} {\bibfnamefont
  {A.}~\bibnamefont {{Brazier}}}, \bibinfo {author} {\bibfnamefont
  {S.}~\bibnamefont {{Bhatnagar}}}, \bibinfo {author} {\bibfnamefont
  {S.}~\bibnamefont {{Myers}}}, \bibinfo {author} {\bibfnamefont
  {F.}~\bibnamefont {{Walter}}},\ and\ \bibinfo {author} {\bibfnamefont
  {B.}~\bibnamefont {{Gaensler}}},\ }in\ \href@noop {} {\emph {\bibinfo
  {booktitle} {Bulletin of the American Astronomical Society}}},\ Vol.~\bibinfo
  {volume} {51}\ (\bibinfo {year} {2019})\ p.\ \bibinfo {pages} {255},\ \Eprint
  {https://arxiv.org/abs/1907.07648} {arXiv:1907.07648 [astro-ph.IM]}
  \BibitemShut {NoStop}%
\bibitem [{Note1()}]{Note1}%
  \BibitemOpen
  \bibinfo {note} {Relatedly, strong lensing of fast radio bursts (FRBs) has
  already been proposed as a powerful probe of massive compact halo objects
  such as primordial black holes in the mass range of $10^{-4}-10^4M_{\odot }$
  \cite {Munoz:2016tmg,Kader:2022jqp}.}\BibitemShut {Stop}%
\bibitem [{\citenamefont {Vachaspati}\ and\ \citenamefont
  {Vilenkin}(1984)}]{Vachaspati:1984dz}%
  \BibitemOpen
  \bibfield  {author} {\bibinfo {author} {\bibfnamefont {T.}~\bibnamefont
  {Vachaspati}}\ and\ \bibinfo {author} {\bibfnamefont {A.}~\bibnamefont
  {Vilenkin}},\ }\href {https://doi.org/10.1103/PhysRevD.30.2036} {\bibfield
  {journal} {\bibinfo  {journal} {Phys. Rev. D}\ }\textbf {\bibinfo {volume}
  {30}},\ \bibinfo {pages} {2036} (\bibinfo {year} {1984})}\BibitemShut
  {NoStop}%
\bibitem [{\citenamefont {Gorghetto}\ \emph {et~al.}(2018)\citenamefont
  {Gorghetto}, \citenamefont {Hardy},\ and\ \citenamefont
  {Villadoro}}]{Gorghetto:2018myk}%
  \BibitemOpen
  \bibfield  {author} {\bibinfo {author} {\bibfnamefont {M.}~\bibnamefont
  {Gorghetto}}, \bibinfo {author} {\bibfnamefont {E.}~\bibnamefont {Hardy}},\
  and\ \bibinfo {author} {\bibfnamefont {G.}~\bibnamefont {Villadoro}},\ }\href
  {https://doi.org/10.1007/JHEP07(2018)151} {\bibfield  {journal} {\bibinfo
  {journal} {JHEP}\ }\textbf {\bibinfo {volume} {07}},\ \bibinfo {pages}
  {151}},\ \Eprint {https://arxiv.org/abs/1806.04677} {arXiv:1806.04677
  [hep-ph]} \BibitemShut {NoStop}%
\bibitem [{\citenamefont {Vilenkin}(1981)}]{PhysRevD.23.852}%
  \BibitemOpen
  \bibfield  {author} {\bibinfo {author} {\bibfnamefont {A.}~\bibnamefont
  {Vilenkin}},\ }\href {https://doi.org/10.1103/PhysRevD.23.852} {\bibfield
  {journal} {\bibinfo  {journal} {Phys. Rev. D}\ }\textbf {\bibinfo {volume}
  {23}},\ \bibinfo {pages} {852} (\bibinfo {year} {1981})}\BibitemShut
  {NoStop}%
\bibitem [{\citenamefont {{Vilenkin}}(1984)}]{1984ApJ...282L..51V}%
  \BibitemOpen
  \bibfield  {author} {\bibinfo {author} {\bibfnamefont {A.}~\bibnamefont
  {{Vilenkin}}},\ }\href {https://doi.org/10.1086/184303} {\bibfield  {journal}
  {\bibinfo  {journal} {\apjl}\ }\textbf {\bibinfo {volume} {282}},\ \bibinfo
  {pages} {L51} (\bibinfo {year} {1984})}\BibitemShut {NoStop}%
\bibitem [{\citenamefont {Almeida}\ and\ \citenamefont
  {Martins}(2021)}]{Almeida:2021ihc}%
  \BibitemOpen
  \bibfield  {author} {\bibinfo {author} {\bibfnamefont {A.~R.~R.}\
  \bibnamefont {Almeida}}\ and\ \bibinfo {author} {\bibfnamefont {C.~J. A.~P.}\
  \bibnamefont {Martins}},\ }\href
  {https://doi.org/10.1103/PhysRevD.104.043524} {\bibfield  {journal} {\bibinfo
   {journal} {Phys. Rev. D}\ }\textbf {\bibinfo {volume} {104}},\ \bibinfo
  {pages} {043524} (\bibinfo {year} {2021})},\ \Eprint
  {https://arxiv.org/abs/2107.11653} {arXiv:2107.11653 [astro-ph.CO]}
  \BibitemShut {NoStop}%
\bibitem [{\citenamefont {Buschmann}\ \emph {et~al.}(2022)\citenamefont
  {Buschmann}, \citenamefont {Foster}, \citenamefont {Hook}, \citenamefont
  {Peterson}, \citenamefont {Willcox}, \citenamefont {Zhang},\ and\
  \citenamefont {Safdi}}]{Buschmann:2021sdq}%
  \BibitemOpen
  \bibfield  {author} {\bibinfo {author} {\bibfnamefont {M.}~\bibnamefont
  {Buschmann}}, \bibinfo {author} {\bibfnamefont {J.~W.}\ \bibnamefont
  {Foster}}, \bibinfo {author} {\bibfnamefont {A.}~\bibnamefont {Hook}},
  \bibinfo {author} {\bibfnamefont {A.}~\bibnamefont {Peterson}}, \bibinfo
  {author} {\bibfnamefont {D.~E.}\ \bibnamefont {Willcox}}, \bibinfo {author}
  {\bibfnamefont {W.}~\bibnamefont {Zhang}},\ and\ \bibinfo {author}
  {\bibfnamefont {B.~R.}\ \bibnamefont {Safdi}},\ }\href
  {https://doi.org/10.1038/s41467-022-28669-y} {\bibfield  {journal} {\bibinfo
  {journal} {Nature Commun.}\ }\textbf {\bibinfo {volume} {13}},\ \bibinfo
  {pages} {1049} (\bibinfo {year} {2022})},\ \Eprint
  {https://arxiv.org/abs/2108.05368} {arXiv:2108.05368 [hep-ph]} \BibitemShut
  {NoStop}%
\bibitem [{Note2()}]{Note2}%
  \BibitemOpen
  \bibinfo {note} {Axion strings arise from a broken global symmetry and hence
  have a more extended energy density profile than gauge strings do, which may
  affect the above scaling.}\BibitemShut {Stop}%
\bibitem [{\citenamefont {Allen}\ and\ \citenamefont
  {Shellard}(1990)}]{Allen:1990tv}%
  \BibitemOpen
  \bibfield  {author} {\bibinfo {author} {\bibfnamefont {B.}~\bibnamefont
  {Allen}}\ and\ \bibinfo {author} {\bibfnamefont {E.~P.~S.}\ \bibnamefont
  {Shellard}},\ }\href {https://doi.org/10.1103/PhysRevLett.64.119} {\bibfield
  {journal} {\bibinfo  {journal} {Phys. Rev. Lett.}\ }\textbf {\bibinfo
  {volume} {64}},\ \bibinfo {pages} {119} (\bibinfo {year} {1990})}\BibitemShut
  {NoStop}%
\bibitem [{\citenamefont {Blanco-Pillado}\ \emph
  {et~al.}(2011{\natexlab{b}})\citenamefont {Blanco-Pillado}, \citenamefont
  {Olum},\ and\ \citenamefont {Shlaer}}]{Blanco-Pillado:2011egf}%
  \BibitemOpen
  \bibfield  {author} {\bibinfo {author} {\bibfnamefont {J.~J.}\ \bibnamefont
  {Blanco-Pillado}}, \bibinfo {author} {\bibfnamefont {K.~D.}\ \bibnamefont
  {Olum}},\ and\ \bibinfo {author} {\bibfnamefont {B.}~\bibnamefont {Shlaer}},\
  }\href {https://doi.org/10.1103/PhysRevD.83.083514} {\bibfield  {journal}
  {\bibinfo  {journal} {Phys. Rev. D}\ }\textbf {\bibinfo {volume} {83}},\
  \bibinfo {pages} {083514} (\bibinfo {year} {2011}{\natexlab{b}})},\ \Eprint
  {https://arxiv.org/abs/1101.5173} {arXiv:1101.5173 [astro-ph.CO]}
  \BibitemShut {NoStop}%
\bibitem [{\citenamefont {Andersen}\ \emph {et~al.}(2020)\citenamefont
  {Andersen} \emph {et~al.}}]{CHIMEFRB:2020abu}%
  \BibitemOpen
  \bibfield  {author} {\bibinfo {author} {\bibfnamefont {B.~C.}\ \bibnamefont
  {Andersen}} \emph {et~al.} (\bibinfo {collaboration} {CHIME/FRB}),\ }\href
  {https://doi.org/10.1038/s41586-020-2863-y} {\bibfield  {journal} {\bibinfo
  {journal} {Nature}\ }\textbf {\bibinfo {volume} {587}},\ \bibinfo {pages}
  {54} (\bibinfo {year} {2020})},\ \Eprint {https://arxiv.org/abs/2005.10324}
  {arXiv:2005.10324 [astro-ph.HE]} \BibitemShut {NoStop}%
\bibitem [{\citenamefont {{Bochenek}}\ \emph {et~al.}(2020)\citenamefont
  {{Bochenek}}, \citenamefont {{Ravi}}, \citenamefont {{Belov}}, \citenamefont
  {{Hallinan}}, \citenamefont {{Kocz}}, \citenamefont {{Kulkarni}},\ and\
  \citenamefont {{McKenna}}}]{2020Natur.587...59B}%
  \BibitemOpen
  \bibfield  {author} {\bibinfo {author} {\bibfnamefont {C.~D.}\ \bibnamefont
  {{Bochenek}}}, \bibinfo {author} {\bibfnamefont {V.}~\bibnamefont {{Ravi}}},
  \bibinfo {author} {\bibfnamefont {K.~V.}\ \bibnamefont {{Belov}}}, \bibinfo
  {author} {\bibfnamefont {G.}~\bibnamefont {{Hallinan}}}, \bibinfo {author}
  {\bibfnamefont {J.}~\bibnamefont {{Kocz}}}, \bibinfo {author} {\bibfnamefont
  {S.~R.}\ \bibnamefont {{Kulkarni}}},\ and\ \bibinfo {author} {\bibfnamefont
  {D.~L.}\ \bibnamefont {{McKenna}}},\ }\href
  {https://doi.org/10.1038/s41586-020-2872-x} {\bibfield  {journal} {\bibinfo
  {journal} {\nat}\ }\textbf {\bibinfo {volume} {587}},\ \bibinfo {pages} {59}
  (\bibinfo {year} {2020})},\ \Eprint {https://arxiv.org/abs/2005.10828}
  {arXiv:2005.10828 [astro-ph.HE]} \BibitemShut {NoStop}%
\bibitem [{\citenamefont {Li}\ and\ \citenamefont {Zhang}(2020)}]{Li:2020esc}%
  \BibitemOpen
  \bibfield  {author} {\bibinfo {author} {\bibfnamefont {Y.}~\bibnamefont
  {Li}}\ and\ \bibinfo {author} {\bibfnamefont {B.}~\bibnamefont {Zhang}},\
  }\href {https://doi.org/10.3847/2041-8213/aba907} {\bibfield  {journal}
  {\bibinfo  {journal} {Astrophys. J. Lett.}\ }\textbf {\bibinfo {volume}
  {899}},\ \bibinfo {pages} {L6} (\bibinfo {year} {2020})},\ \Eprint
  {https://arxiv.org/abs/2005.02371} {arXiv:2005.02371 [astro-ph.HE]}
  \BibitemShut {NoStop}%
\bibitem [{\citenamefont {{Bhardwaj}}\ \emph {et~al.}(2021)\citenamefont
  {{Bhardwaj}}, \citenamefont {{Gaensler}}, \citenamefont {{Kaspi}},
  \citenamefont {{Landecker}}, \citenamefont {{Mckinven}}, \citenamefont
  {{Michilli}}, \citenamefont {{Pleunis}}, \citenamefont {{Tendulkar}},
  \citenamefont {{Andersen}}, \citenamefont {{Boyle}}, \citenamefont
  {{Cassanelli}}, \citenamefont {{Chawla}}, \citenamefont {{Cook}},
  \citenamefont {{Dobbs}}, \citenamefont {{Fonseca}}, \citenamefont
  {{Kaczmarek}}, \citenamefont {{Leung}}, \citenamefont {{Masui}},
  \citenamefont {{Mnchmeyer}}, \citenamefont {{Ng}}, \citenamefont
  {{Rafiei-Ravandi}}, \citenamefont {{Scholz}}, \citenamefont {{Shin}},
  \citenamefont {{Smith}}, \citenamefont {{Stairs}},\ and\ \citenamefont
  {{Zwaniga}}}]{2021ApJ...910L..18B}%
  \BibitemOpen
  \bibfield  {author} {\bibinfo {author} {\bibfnamefont {M.}~\bibnamefont
  {{Bhardwaj}}}, \bibinfo {author} {\bibfnamefont {B.~M.}\ \bibnamefont
  {{Gaensler}}}, \bibinfo {author} {\bibfnamefont {V.~M.}\ \bibnamefont
  {{Kaspi}}}, \bibinfo {author} {\bibfnamefont {T.~L.}\ \bibnamefont
  {{Landecker}}}, \bibinfo {author} {\bibfnamefont {R.}~\bibnamefont
  {{Mckinven}}}, \bibinfo {author} {\bibfnamefont {D.}~\bibnamefont
  {{Michilli}}}, \bibinfo {author} {\bibfnamefont {Z.}~\bibnamefont
  {{Pleunis}}}, \bibinfo {author} {\bibfnamefont {S.~P.}\ \bibnamefont
  {{Tendulkar}}}, \bibinfo {author} {\bibfnamefont {B.~C.}\ \bibnamefont
  {{Andersen}}}, \bibinfo {author} {\bibfnamefont {P.~J.}\ \bibnamefont
  {{Boyle}}}, \bibinfo {author} {\bibfnamefont {T.}~\bibnamefont
  {{Cassanelli}}}, \bibinfo {author} {\bibfnamefont {P.}~\bibnamefont
  {{Chawla}}}, \bibinfo {author} {\bibfnamefont {A.}~\bibnamefont {{Cook}}},
  \bibinfo {author} {\bibfnamefont {M.}~\bibnamefont {{Dobbs}}}, \bibinfo
  {author} {\bibfnamefont {E.}~\bibnamefont {{Fonseca}}}, \bibinfo {author}
  {\bibfnamefont {J.}~\bibnamefont {{Kaczmarek}}}, \bibinfo {author}
  {\bibfnamefont {C.}~\bibnamefont {{Leung}}}, \bibinfo {author} {\bibfnamefont
  {K.}~\bibnamefont {{Masui}}}, \bibinfo {author} {\bibfnamefont
  {M.}~\bibnamefont {{Mnchmeyer}}}, \bibinfo {author} {\bibfnamefont
  {C.}~\bibnamefont {{Ng}}}, \bibinfo {author} {\bibfnamefont {M.}~\bibnamefont
  {{Rafiei-Ravandi}}}, \bibinfo {author} {\bibfnamefont {P.}~\bibnamefont
  {{Scholz}}}, \bibinfo {author} {\bibfnamefont {K.}~\bibnamefont {{Shin}}},
  \bibinfo {author} {\bibfnamefont {K.~M.}\ \bibnamefont {{Smith}}}, \bibinfo
  {author} {\bibfnamefont {I.~H.}\ \bibnamefont {{Stairs}}},\ and\ \bibinfo
  {author} {\bibfnamefont {A.~V.}\ \bibnamefont {{Zwaniga}}},\ }\href
  {https://doi.org/10.3847/2041-8213/abeaa6} {\bibfield  {journal} {\bibinfo
  {journal} {\apjl}\ }\textbf {\bibinfo {volume} {910}},\ \bibinfo {eid} {L18}
  (\bibinfo {year} {2021})},\ \Eprint {https://arxiv.org/abs/2103.01295}
  {arXiv:2103.01295 [astro-ph.HE]} \BibitemShut {NoStop}%
\bibitem [{\citenamefont {Kirsten}\ \emph {et~al.}(2022)\citenamefont {Kirsten}
  \emph {et~al.}}]{Kirsten:2021llv}%
  \BibitemOpen
  \bibfield  {author} {\bibinfo {author} {\bibfnamefont {F.}~\bibnamefont
  {Kirsten}} \emph {et~al.},\ }\href
  {https://doi.org/10.1038/s41586-021-04354-w} {\bibfield  {journal} {\bibinfo
  {journal} {Nature}\ }\textbf {\bibinfo {volume} {602}},\ \bibinfo {pages}
  {585} (\bibinfo {year} {2022})},\ \Eprint {https://arxiv.org/abs/2105.11445}
  {arXiv:2105.11445 [astro-ph.HE]} \BibitemShut {NoStop}%
\bibitem [{\citenamefont {Luo}\ \emph {et~al.}(2018)\citenamefont {Luo},
  \citenamefont {Lee}, \citenamefont {Lorimer},\ and\ \citenamefont
  {Zhang}}]{Luo:2018tiy}%
  \BibitemOpen
  \bibfield  {author} {\bibinfo {author} {\bibfnamefont {R.}~\bibnamefont
  {Luo}}, \bibinfo {author} {\bibfnamefont {K.}~\bibnamefont {Lee}}, \bibinfo
  {author} {\bibfnamefont {D.~R.}\ \bibnamefont {Lorimer}},\ and\ \bibinfo
  {author} {\bibfnamefont {B.}~\bibnamefont {Zhang}},\ }\href
  {https://doi.org/10.1093/mnras/sty2364} {\bibfield  {journal} {\bibinfo
  {journal} {Mon. Not. Roy. Astron. Soc.}\ }\textbf {\bibinfo {volume} {481}},\
  \bibinfo {pages} {2320} (\bibinfo {year} {2018})},\ \Eprint
  {https://arxiv.org/abs/1808.09929} {arXiv:1808.09929 [astro-ph.HE]}
  \BibitemShut {NoStop}%
\bibitem [{\citenamefont {Luo}\ \emph {et~al.}(2020)\citenamefont {Luo},
  \citenamefont {Men}, \citenamefont {Lee}, \citenamefont {Wang}, \citenamefont
  {Lorimer},\ and\ \citenamefont {Zhang}}]{Luo:2020wfx}%
  \BibitemOpen
  \bibfield  {author} {\bibinfo {author} {\bibfnamefont {R.}~\bibnamefont
  {Luo}}, \bibinfo {author} {\bibfnamefont {Y.}~\bibnamefont {Men}}, \bibinfo
  {author} {\bibfnamefont {K.}~\bibnamefont {Lee}}, \bibinfo {author}
  {\bibfnamefont {W.}~\bibnamefont {Wang}}, \bibinfo {author} {\bibfnamefont
  {D.~R.}\ \bibnamefont {Lorimer}},\ and\ \bibinfo {author} {\bibfnamefont
  {B.}~\bibnamefont {Zhang}},\ }\href {https://doi.org/10.1093/mnras/staa704}
  {\bibfield  {journal} {\bibinfo  {journal} {Mon. Not. Roy. Astron. Soc.}\
  }\textbf {\bibinfo {volume} {494}},\ \bibinfo {pages} {665} (\bibinfo {year}
  {2020})},\ \Eprint {https://arxiv.org/abs/2003.04848} {arXiv:2003.04848
  [astro-ph.HE]} \BibitemShut {NoStop}%
\bibitem [{\citenamefont {{Schechter}}(1976)}]{1976ApJ...203..297S}%
  \BibitemOpen
  \bibfield  {author} {\bibinfo {author} {\bibfnamefont {P.}~\bibnamefont
  {{Schechter}}},\ }\href {https://doi.org/10.1086/154079} {\bibfield
  {journal} {\bibinfo  {journal} {\apj}\ }\textbf {\bibinfo {volume} {203}},\
  \bibinfo {pages} {297} (\bibinfo {year} {1976})}\BibitemShut {NoStop}%
\bibitem [{\citenamefont {Zhang}\ \emph {et~al.}(2021)\citenamefont {Zhang},
  \citenamefont {Zhang}, \citenamefont {Li},\ and\ \citenamefont
  {Lorimer}}]{Zhang:2020ass}%
  \BibitemOpen
  \bibfield  {author} {\bibinfo {author} {\bibfnamefont {R.~C.}\ \bibnamefont
  {Zhang}}, \bibinfo {author} {\bibfnamefont {B.}~\bibnamefont {Zhang}},
  \bibinfo {author} {\bibfnamefont {Y.}~\bibnamefont {Li}},\ and\ \bibinfo
  {author} {\bibfnamefont {D.~R.}\ \bibnamefont {Lorimer}},\ }\href
  {https://doi.org/10.1093/mnras/staa3537} {\bibfield  {journal} {\bibinfo
  {journal} {Mon. Not. Roy. Astron. Soc.}\ }\textbf {\bibinfo {volume} {501}},\
  \bibinfo {pages} {157} (\bibinfo {year} {2021})},\ \Eprint
  {https://arxiv.org/abs/2011.06151} {arXiv:2011.06151 [astro-ph.HE]}
  \BibitemShut {NoStop}%
\bibitem [{\citenamefont {Zhang}\ and\ \citenamefont
  {Zhang}(2022)}]{Zhang:2021kdu}%
  \BibitemOpen
  \bibfield  {author} {\bibinfo {author} {\bibfnamefont {R.~C.}\ \bibnamefont
  {Zhang}}\ and\ \bibinfo {author} {\bibfnamefont {B.}~\bibnamefont {Zhang}},\
  }\href {https://doi.org/10.3847/2041-8213/ac46ad} {\bibfield  {journal}
  {\bibinfo  {journal} {Astrophys. J. Lett.}\ }\textbf {\bibinfo {volume}
  {924}},\ \bibinfo {pages} {L14} (\bibinfo {year} {2022})},\ \Eprint
  {https://arxiv.org/abs/2109.07558} {arXiv:2109.07558 [astro-ph.HE]}
  \BibitemShut {NoStop}%
\bibitem [{\citenamefont {Qiang}\ \emph {et~al.}(2022)\citenamefont {Qiang},
  \citenamefont {Li},\ and\ \citenamefont {Wei}}]{Qiang:2021ljr}%
  \BibitemOpen
  \bibfield  {author} {\bibinfo {author} {\bibfnamefont {D.-C.}\ \bibnamefont
  {Qiang}}, \bibinfo {author} {\bibfnamefont {S.-L.}\ \bibnamefont {Li}},\ and\
  \bibinfo {author} {\bibfnamefont {H.}~\bibnamefont {Wei}},\ }\href
  {https://doi.org/10.1088/1475-7516/2022/01/040} {\bibfield  {journal}
  {\bibinfo  {journal} {JCAP}\ }\textbf {\bibinfo {volume} {01}},\ \bibinfo
  {pages} {040}},\ \Eprint {https://arxiv.org/abs/2111.07476} {arXiv:2111.07476
  [astro-ph.HE]} \BibitemShut {NoStop}%
\bibitem [{\citenamefont {Macquart}\ \emph
  {et~al.}(2019{\natexlab{a}})\citenamefont {Macquart}, \citenamefont
  {Shannon}, \citenamefont {Bannister}, \citenamefont {James}, \citenamefont
  {Ekers},\ and\ \citenamefont {Bunton}}]{2019}%
  \BibitemOpen
  \bibfield  {author} {\bibinfo {author} {\bibfnamefont {J.-P.}\ \bibnamefont
  {Macquart}}, \bibinfo {author} {\bibfnamefont {R.~M.}\ \bibnamefont
  {Shannon}}, \bibinfo {author} {\bibfnamefont {K.~W.}\ \bibnamefont
  {Bannister}}, \bibinfo {author} {\bibfnamefont {C.~W.}\ \bibnamefont
  {James}}, \bibinfo {author} {\bibfnamefont {R.~D.}\ \bibnamefont {Ekers}},\
  and\ \bibinfo {author} {\bibfnamefont {J.~D.}\ \bibnamefont {Bunton}},\
  }\href {https://doi.org/10.3847/2041-8213/ab03d6} {\bibfield  {journal}
  {\bibinfo  {journal} {The Astrophysical Journal}\ }\textbf {\bibinfo {volume}
  {872}},\ \bibinfo {pages} {L19} (\bibinfo {year}
  {2019}{\natexlab{a}})}\BibitemShut {NoStop}%
\bibitem [{\citenamefont {Petroff}\ \emph {et~al.}(2016)\citenamefont
  {Petroff}, \citenamefont {Barr}, \citenamefont {Jameson}, \citenamefont
  {Keane}, \citenamefont {Bailes}, \citenamefont {Kramer}, \citenamefont
  {Morello}, \citenamefont {Tabbara},\ and\ \citenamefont {van
  Straten}}]{Petroff:2016tcr}%
  \BibitemOpen
  \bibfield  {author} {\bibinfo {author} {\bibfnamefont {E.}~\bibnamefont
  {Petroff}}, \bibinfo {author} {\bibfnamefont {E.~D.}\ \bibnamefont {Barr}},
  \bibinfo {author} {\bibfnamefont {A.}~\bibnamefont {Jameson}}, \bibinfo
  {author} {\bibfnamefont {E.~F.}\ \bibnamefont {Keane}}, \bibinfo {author}
  {\bibfnamefont {M.}~\bibnamefont {Bailes}}, \bibinfo {author} {\bibfnamefont
  {M.}~\bibnamefont {Kramer}}, \bibinfo {author} {\bibfnamefont
  {V.}~\bibnamefont {Morello}}, \bibinfo {author} {\bibfnamefont
  {D.}~\bibnamefont {Tabbara}},\ and\ \bibinfo {author} {\bibfnamefont
  {W.}~\bibnamefont {van Straten}},\ }\href
  {https://doi.org/10.1017/pasa.2016.35} {\bibfield  {journal} {\bibinfo
  {journal} {Publ. Astron. Soc. Austral.}\ }\textbf {\bibinfo {volume} {33}},\
  \bibinfo {pages} {e045} (\bibinfo {year} {2016})},\ \Eprint
  {https://arxiv.org/abs/1601.03547} {arXiv:1601.03547 [astro-ph.HE]}
  \BibitemShut {NoStop}%
\bibitem [{\citenamefont {{Gott}}(1985)}]{1985ApJ...288..422G}%
  \BibitemOpen
  \bibfield  {author} {\bibinfo {author} {\bibfnamefont {I.}~\bibnamefont
  {{Gott}}, \bibfnamefont {J.~R.}},\ }\href {https://doi.org/10.1086/162808}
  {\bibfield  {journal} {\bibinfo  {journal} {\apj}\ }\textbf {\bibinfo
  {volume} {288}},\ \bibinfo {pages} {422} (\bibinfo {year}
  {1985})}\BibitemShut {NoStop}%
\bibitem [{\citenamefont {Agrawal}\ \emph {et~al.}(2022)\citenamefont
  {Agrawal}, \citenamefont {Hook}, \citenamefont {Huang},\ and\ \citenamefont
  {Marques-Tavares}}]{Agrawal:2020euj}%
  \BibitemOpen
  \bibfield  {author} {\bibinfo {author} {\bibfnamefont {P.}~\bibnamefont
  {Agrawal}}, \bibinfo {author} {\bibfnamefont {A.}~\bibnamefont {Hook}},
  \bibinfo {author} {\bibfnamefont {J.}~\bibnamefont {Huang}},\ and\ \bibinfo
  {author} {\bibfnamefont {G.}~\bibnamefont {Marques-Tavares}},\ }\href
  {https://doi.org/10.1007/JHEP01(2022)103} {\bibfield  {journal} {\bibinfo
  {journal} {JHEP}\ }\textbf {\bibinfo {volume} {01}},\ \bibinfo {pages}
  {103}},\ \Eprint {https://arxiv.org/abs/2010.15848} {arXiv:2010.15848
  [hep-ph]} \BibitemShut {NoStop}%
\bibitem [{\citenamefont {Dai}\ and\ \citenamefont {Lu}(2017)}]{Dai:2017twh}%
  \BibitemOpen
  \bibfield  {author} {\bibinfo {author} {\bibfnamefont {L.}~\bibnamefont
  {Dai}}\ and\ \bibinfo {author} {\bibfnamefont {W.}~\bibnamefont {Lu}},\
  }\href {https://doi.org/10.3847/1538-4357/aa8873} {\bibfield  {journal}
  {\bibinfo  {journal} {Astrophys. J.}\ }\textbf {\bibinfo {volume} {847}},\
  \bibinfo {pages} {19} (\bibinfo {year} {2017})},\ \Eprint
  {https://arxiv.org/abs/1706.06103} {arXiv:1706.06103 [astro-ph.HE]}
  \BibitemShut {NoStop}%
\bibitem [{\citenamefont {{Pearson}}\ \emph {et~al.}(2021)\citenamefont
  {{Pearson}}, \citenamefont {{Trendafilova}},\ and\ \citenamefont
  {{Meyers}}}]{Pearson2021lensedFRBgw}%
  \BibitemOpen
  \bibfield  {author} {\bibinfo {author} {\bibfnamefont {N.}~\bibnamefont
  {{Pearson}}}, \bibinfo {author} {\bibfnamefont {C.}~\bibnamefont
  {{Trendafilova}}},\ and\ \bibinfo {author} {\bibfnamefont {J.}~\bibnamefont
  {{Meyers}}},\ }\href {https://doi.org/10.1103/PhysRevD.103.063017} {\bibfield
   {journal} {\bibinfo  {journal} {\prd}\ }\textbf {\bibinfo {volume} {103}},\
  \bibinfo {eid} {063017} (\bibinfo {year} {2021})},\ \Eprint
  {https://arxiv.org/abs/2009.11252} {arXiv:2009.11252 [astro-ph.CO]}
  \BibitemShut {NoStop}%
\bibitem [{\citenamefont {{Wucknitz}}\ \emph {et~al.}(2021)\citenamefont
  {{Wucknitz}}, \citenamefont {{Spitler}},\ and\ \citenamefont
  {{Pen}}}]{Wucknitz2021lensedFRBcosmology}%
  \BibitemOpen
  \bibfield  {author} {\bibinfo {author} {\bibfnamefont {O.}~\bibnamefont
  {{Wucknitz}}}, \bibinfo {author} {\bibfnamefont {L.~G.}\ \bibnamefont
  {{Spitler}}},\ and\ \bibinfo {author} {\bibfnamefont {U.~L.}\ \bibnamefont
  {{Pen}}},\ }\href {https://doi.org/10.1051/0004-6361/202038248} {\bibfield
  {journal} {\bibinfo  {journal} {\aap}\ }\textbf {\bibinfo {volume} {645}},\
  \bibinfo {eid} {A44} (\bibinfo {year} {2021})},\ \Eprint
  {https://arxiv.org/abs/2004.11643} {arXiv:2004.11643 [astro-ph.CO]}
  \BibitemShut {NoStop}%
\bibitem [{\citenamefont {Petroff}\ \emph {et~al.}(2019)\citenamefont
  {Petroff}, \citenamefont {Hessels},\ and\ \citenamefont
  {Lorimer}}]{Petroff:2019tty}%
  \BibitemOpen
  \bibfield  {author} {\bibinfo {author} {\bibfnamefont {E.}~\bibnamefont
  {Petroff}}, \bibinfo {author} {\bibfnamefont {J.~W.~T.}\ \bibnamefont
  {Hessels}},\ and\ \bibinfo {author} {\bibfnamefont {D.~R.}\ \bibnamefont
  {Lorimer}},\ }\href {https://doi.org/10.1007/s00159-019-0116-6} {\bibfield
  {journal} {\bibinfo  {journal} {Astron. Astrophys. Rev.}\ }\textbf {\bibinfo
  {volume} {27}},\ \bibinfo {pages} {4} (\bibinfo {year} {2019})},\ \Eprint
  {https://arxiv.org/abs/1904.07947} {arXiv:1904.07947 [astro-ph.HE]}
  \BibitemShut {NoStop}%
\bibitem [{\citenamefont {Amiri}\ \emph {et~al.}(2021)\citenamefont {Amiri}
  \emph {et~al.}}]{CHIMEFRB:2021srp}%
  \BibitemOpen
  \bibfield  {author} {\bibinfo {author} {\bibfnamefont {M.}~\bibnamefont
  {Amiri}} \emph {et~al.} (\bibinfo {collaboration} {CHIME/FRB}),\ }\href
  {https://doi.org/10.3847/1538-4365/ac33ab} {\bibfield  {journal} {\bibinfo
  {journal} {Astrophys. J. Supp.}\ }\textbf {\bibinfo {volume} {257}},\
  \bibinfo {pages} {59} (\bibinfo {year} {2021})},\ \Eprint
  {https://arxiv.org/abs/2106.04352} {arXiv:2106.04352 [astro-ph.HE]}
  \BibitemShut {NoStop}%
\bibitem [{\citenamefont {Lin}\ \emph {et~al.}(2022)\citenamefont {Lin} \emph
  {et~al.}}]{Lin:2022wgp}%
  \BibitemOpen
  \bibfield  {author} {\bibinfo {author} {\bibfnamefont {H.-H.}\ \bibnamefont
  {Lin}} \emph {et~al.},\ }\href@noop {} {\  (\bibinfo {year} {2022})},\
  \Eprint {https://arxiv.org/abs/2206.08983} {arXiv:2206.08983 [astro-ph.IM]}
  \BibitemShut {NoStop}%
\bibitem [{\citenamefont {Slosar}\ \emph {et~al.}(2019)\citenamefont {Slosar}
  \emph {et~al.}}]{PUMA:2019jwd}%
  \BibitemOpen
  \bibfield  {author} {\bibinfo {author} {\bibfnamefont {A.}~\bibnamefont
  {Slosar}} \emph {et~al.} (\bibinfo {collaboration} {PUMA}),\ }\href@noop {}
  {\bibfield  {journal} {\bibinfo  {journal} {Bull. Am. Astron. Soc.}\ }\textbf
  {\bibinfo {volume} {51}},\ \bibinfo {pages} {53} (\bibinfo {year} {2019})},\
  \Eprint {https://arxiv.org/abs/1907.12559} {arXiv:1907.12559 [astro-ph.IM]}
  \BibitemShut {NoStop}%
\bibitem [{\citenamefont {Macquart}\ \emph
  {et~al.}(2019{\natexlab{b}})\citenamefont {Macquart}, \citenamefont
  {Shannon}, \citenamefont {Bannister}, \citenamefont {James}, \citenamefont
  {Ekers},\ and\ \citenamefont {Bunton}}]{Macquart_2019}%
  \BibitemOpen
  \bibfield  {author} {\bibinfo {author} {\bibfnamefont {J.-P.}\ \bibnamefont
  {Macquart}}, \bibinfo {author} {\bibfnamefont {R.~M.}\ \bibnamefont
  {Shannon}}, \bibinfo {author} {\bibfnamefont {K.~W.}\ \bibnamefont
  {Bannister}}, \bibinfo {author} {\bibfnamefont {C.~W.}\ \bibnamefont
  {James}}, \bibinfo {author} {\bibfnamefont {R.~D.}\ \bibnamefont {Ekers}},\
  and\ \bibinfo {author} {\bibfnamefont {J.~D.}\ \bibnamefont {Bunton}},\
  }\href {https://doi.org/10.3847/2041-8213/ab03d6} {\bibfield  {journal}
  {\bibinfo  {journal} {The Astrophysical Journal}\ }\textbf {\bibinfo {volume}
  {872}},\ \bibinfo {pages} {L19} (\bibinfo {year}
  {2019}{\natexlab{b}})}\BibitemShut {NoStop}%
\bibitem [{\citenamefont {Metzger}\ \emph {et~al.}(2019)\citenamefont
  {Metzger}, \citenamefont {Margalit},\ and\ \citenamefont
  {Sironi}}]{Metzger:2019una}%
  \BibitemOpen
  \bibfield  {author} {\bibinfo {author} {\bibfnamefont {B.~D.}\ \bibnamefont
  {Metzger}}, \bibinfo {author} {\bibfnamefont {B.}~\bibnamefont {Margalit}},\
  and\ \bibinfo {author} {\bibfnamefont {L.}~\bibnamefont {Sironi}},\ }\href
  {https://doi.org/10.1093/mnras/stz700} {\bibfield  {journal} {\bibinfo
  {journal} {Mon. Not. Roy. Astron. Soc.}\ }\textbf {\bibinfo {volume} {485}},\
  \bibinfo {pages} {4091} (\bibinfo {year} {2019})},\ \Eprint
  {https://arxiv.org/abs/1902.01866} {arXiv:1902.01866 [astro-ph.HE]}
  \BibitemShut {NoStop}%
\bibitem [{Note3()}]{Note3}%
  \BibitemOpen
  \bibinfo {note} {In detail, $\ell < (2 \Gamma ^2 \protect \, c \protect \,
  \delta t_{\protect \rm var})/\Gamma $, larger by a factor of the Lorentz
  factor but still smaller than we require for $\Gamma \lesssim 10^3$ for a
  variability timescale of $\delta t_{\protect \rm var} = 1$ ms, a condition on
  the Lorentz factor that is satisfied for the fiducial parameters taken in
  external shock models. Some bursts suggests $\delta t_{\protect \rm var}
  \lesssim 1 ~\mu $s.}\BibitemShut {Stop}%
\bibitem [{\citenamefont {Smith}\ \emph {et~al.}(2003)\citenamefont {Smith},
  \citenamefont {Peacock}, \citenamefont {Jenkins}, \citenamefont {White},
  \citenamefont {Frenk}, \citenamefont {Pearce}, \citenamefont {Thomas},
  \citenamefont {Efstathiou},\ and\ \citenamefont {Couchman}}]{Smith_2003}%
  \BibitemOpen
  \bibfield  {author} {\bibinfo {author} {\bibfnamefont {R.~E.}\ \bibnamefont
  {Smith}}, \bibinfo {author} {\bibfnamefont {J.~A.}\ \bibnamefont {Peacock}},
  \bibinfo {author} {\bibfnamefont {A.}~\bibnamefont {Jenkins}}, \bibinfo
  {author} {\bibfnamefont {S.~D.~M.}\ \bibnamefont {White}}, \bibinfo {author}
  {\bibfnamefont {C.~S.}\ \bibnamefont {Frenk}}, \bibinfo {author}
  {\bibfnamefont {F.~R.}\ \bibnamefont {Pearce}}, \bibinfo {author}
  {\bibfnamefont {P.~A.}\ \bibnamefont {Thomas}}, \bibinfo {author}
  {\bibfnamefont {G.}~\bibnamefont {Efstathiou}},\ and\ \bibinfo {author}
  {\bibfnamefont {H.~M.~P.}\ \bibnamefont {Couchman}},\ }\href
  {https://doi.org/10.1046/j.1365-8711.2003.06503.x} {\bibfield  {journal}
  {\bibinfo  {journal} {Monthly Notices of the Royal Astronomical Society}\
  }\textbf {\bibinfo {volume} {341}},\ \bibinfo {pages} {1311} (\bibinfo {year}
  {2003})}\BibitemShut {NoStop}%
\bibitem [{\citenamefont {Dai}\ \emph {et~al.}(2017)\citenamefont {Dai},
  \citenamefont {Venumadhav},\ and\ \citenamefont {Sigurdson}}]{Dai:2016igl}%
  \BibitemOpen
  \bibfield  {author} {\bibinfo {author} {\bibfnamefont {L.}~\bibnamefont
  {Dai}}, \bibinfo {author} {\bibfnamefont {T.}~\bibnamefont {Venumadhav}},\
  and\ \bibinfo {author} {\bibfnamefont {K.}~\bibnamefont {Sigurdson}},\ }\href
  {https://doi.org/10.1103/PhysRevD.95.044011} {\bibfield  {journal} {\bibinfo
  {journal} {Phys. Rev. D}\ }\textbf {\bibinfo {volume} {95}},\ \bibinfo
  {pages} {044011} (\bibinfo {year} {2017})},\ \Eprint
  {https://arxiv.org/abs/1605.09398} {arXiv:1605.09398 [astro-ph.CO]}
  \BibitemShut {NoStop}%
\bibitem [{\citenamefont {Hilbert}\ \emph {et~al.}(2007)\citenamefont
  {Hilbert}, \citenamefont {White}, \citenamefont {Hartlap},\ and\
  \citenamefont {Schneider}}]{Hilbert:2007ny}%
  \BibitemOpen
  \bibfield  {author} {\bibinfo {author} {\bibfnamefont {S.}~\bibnamefont
  {Hilbert}}, \bibinfo {author} {\bibfnamefont {S.~D.~M.}\ \bibnamefont
  {White}}, \bibinfo {author} {\bibfnamefont {J.}~\bibnamefont {Hartlap}},\
  and\ \bibinfo {author} {\bibfnamefont {P.}~\bibnamefont {Schneider}},\ }\href
  {https://doi.org/10.1111/j.1365-2966.2007.12391.x} {\bibfield  {journal}
  {\bibinfo  {journal} {Mon. Not. Roy. Astron. Soc.}\ }\textbf {\bibinfo
  {volume} {382}},\ \bibinfo {pages} {121} (\bibinfo {year} {2007})},\ \Eprint
  {https://arxiv.org/abs/astro-ph/0703803} {arXiv:astro-ph/0703803}
  \BibitemShut {NoStop}%
\bibitem [{\citenamefont {Hilbert}\ \emph {et~al.}(2008)\citenamefont
  {Hilbert}, \citenamefont {White}, \citenamefont {Hartlap},\ and\
  \citenamefont {Schneider}}]{Hilbert:2007jd}%
  \BibitemOpen
  \bibfield  {author} {\bibinfo {author} {\bibfnamefont {S.}~\bibnamefont
  {Hilbert}}, \bibinfo {author} {\bibfnamefont {S.~D.~M.}\ \bibnamefont
  {White}}, \bibinfo {author} {\bibfnamefont {J.}~\bibnamefont {Hartlap}},\
  and\ \bibinfo {author} {\bibfnamefont {P.}~\bibnamefont {Schneider}},\ }\href
  {https://doi.org/10.1111/j.1365-2966.2008.13190.x} {\bibfield  {journal}
  {\bibinfo  {journal} {Mon. Not. Roy. Astron. Soc.}\ }\textbf {\bibinfo
  {volume} {386}},\ \bibinfo {pages} {1845} (\bibinfo {year} {2008})},\ \Eprint
  {https://arxiv.org/abs/0712.1593} {arXiv:0712.1593 [astro-ph]} \BibitemShut
  {NoStop}%
\bibitem [{\citenamefont {Oguri}(2018)}]{Oguri:2018muv}%
  \BibitemOpen
  \bibfield  {author} {\bibinfo {author} {\bibfnamefont {M.}~\bibnamefont
  {Oguri}},\ }\href {https://doi.org/10.1093/mnras/sty2145} {\bibfield
  {journal} {\bibinfo  {journal} {Mon. Not. Roy. Astron. Soc.}\ }\textbf
  {\bibinfo {volume} {480}},\ \bibinfo {pages} {3842} (\bibinfo {year}
  {2018})},\ \Eprint {https://arxiv.org/abs/1807.02584} {arXiv:1807.02584
  [astro-ph.CO]} \BibitemShut {NoStop}%
\bibitem [{\citenamefont {Dai}\ \emph {et~al.}(2018)\citenamefont {Dai},
  \citenamefont {Venumadhav}, \citenamefont {Kaurov},\ and\ \citenamefont
  {Miralda-Escud\'e}}]{Dai:2018mxx}%
  \BibitemOpen
  \bibfield  {author} {\bibinfo {author} {\bibfnamefont {L.}~\bibnamefont
  {Dai}}, \bibinfo {author} {\bibfnamefont {T.}~\bibnamefont {Venumadhav}},
  \bibinfo {author} {\bibfnamefont {A.~A.}\ \bibnamefont {Kaurov}},\ and\
  \bibinfo {author} {\bibfnamefont {J.}~\bibnamefont {Miralda-Escud\'e}},\
  }\href {https://doi.org/10.3847/1538-4357/aae478} {\bibfield  {journal}
  {\bibinfo  {journal} {Astrophys. J.}\ }\textbf {\bibinfo {volume} {867}},\
  \bibinfo {pages} {24} (\bibinfo {year} {2018})},\ \Eprint
  {https://arxiv.org/abs/1804.03149} {arXiv:1804.03149 [astro-ph.CO]}
  \BibitemShut {NoStop}%
\bibitem [{\citenamefont {Dai}\ \emph {et~al.}(2020)\citenamefont {Dai},
  \citenamefont {Kaurov}, \citenamefont {Sharon}, \citenamefont {Florian},
  \citenamefont {Miralda-Escud\'e}, \citenamefont {Venumadhav}, \citenamefont
  {Frye}, \citenamefont {Rigby},\ and\ \citenamefont {Bayliss}}]{Dai:2020rio}%
  \BibitemOpen
  \bibfield  {author} {\bibinfo {author} {\bibfnamefont {L.}~\bibnamefont
  {Dai}}, \bibinfo {author} {\bibfnamefont {A.~A.}\ \bibnamefont {Kaurov}},
  \bibinfo {author} {\bibfnamefont {K.}~\bibnamefont {Sharon}}, \bibinfo
  {author} {\bibfnamefont {M.~K.}\ \bibnamefont {Florian}}, \bibinfo {author}
  {\bibfnamefont {J.}~\bibnamefont {Miralda-Escud\'e}}, \bibinfo {author}
  {\bibfnamefont {T.}~\bibnamefont {Venumadhav}}, \bibinfo {author}
  {\bibfnamefont {B.}~\bibnamefont {Frye}}, \bibinfo {author} {\bibfnamefont
  {J.~R.}\ \bibnamefont {Rigby}},\ and\ \bibinfo {author} {\bibfnamefont
  {M.}~\bibnamefont {Bayliss}},\ }\href
  {https://doi.org/10.1093/mnras/staa1355} {\bibfield  {journal} {\bibinfo
  {journal} {Mon. Not. Roy. Astron. Soc.}\ }\textbf {\bibinfo {volume} {495}},\
  \bibinfo {pages} {3192} (\bibinfo {year} {2020})},\ \Eprint
  {https://arxiv.org/abs/2001.00261} {arXiv:2001.00261 [astro-ph.GA]}
  \BibitemShut {NoStop}%
\bibitem [{\citenamefont {Agrawal}\ \emph {et~al.}(2020)\citenamefont
  {Agrawal}, \citenamefont {Hook},\ and\ \citenamefont
  {Huang}}]{Agrawal:2019lkr}%
  \BibitemOpen
  \bibfield  {author} {\bibinfo {author} {\bibfnamefont {P.}~\bibnamefont
  {Agrawal}}, \bibinfo {author} {\bibfnamefont {A.}~\bibnamefont {Hook}},\ and\
  \bibinfo {author} {\bibfnamefont {J.}~\bibnamefont {Huang}},\ }\href
  {https://doi.org/10.1007/JHEP07(2020)138} {\bibfield  {journal} {\bibinfo
  {journal} {JHEP}\ }\textbf {\bibinfo {volume} {07}},\ \bibinfo {pages}
  {138}},\ \Eprint {https://arxiv.org/abs/1912.02823} {arXiv:1912.02823
  [astro-ph.CO]} \BibitemShut {NoStop}%
\bibitem [{\citenamefont {Jain}\ \emph {et~al.}(2021)\citenamefont {Jain},
  \citenamefont {Long},\ and\ \citenamefont {Amin}}]{Jain:2021shf}%
  \BibitemOpen
  \bibfield  {author} {\bibinfo {author} {\bibfnamefont {M.}~\bibnamefont
  {Jain}}, \bibinfo {author} {\bibfnamefont {A.~J.}\ \bibnamefont {Long}},\
  and\ \bibinfo {author} {\bibfnamefont {M.~A.}\ \bibnamefont {Amin}},\ }\href
  {https://doi.org/10.1088/1475-7516/2021/05/055} {\bibfield  {journal}
  {\bibinfo  {journal} {JCAP}\ }\textbf {\bibinfo {volume} {05}},\ \bibinfo
  {pages} {055}},\ \Eprint {https://arxiv.org/abs/2103.10962} {arXiv:2103.10962
  [astro-ph.CO]} \BibitemShut {NoStop}%
\bibitem [{\citenamefont {Yin}\ \emph {et~al.}(2021)\citenamefont {Yin},
  \citenamefont {Dai},\ and\ \citenamefont {Ferraro}}]{Yin:2021kmx}%
  \BibitemOpen
  \bibfield  {author} {\bibinfo {author} {\bibfnamefont {W.~W.}\ \bibnamefont
  {Yin}}, \bibinfo {author} {\bibfnamefont {L.}~\bibnamefont {Dai}},\ and\
  \bibinfo {author} {\bibfnamefont {S.}~\bibnamefont {Ferraro}},\ }\href@noop
  {} {\  (\bibinfo {year} {2021})},\ \Eprint {https://arxiv.org/abs/2111.12741}
  {arXiv:2111.12741 [astro-ph.CO]} \BibitemShut {NoStop}%
\bibitem [{\citenamefont {Mu\~noz}\ \emph {et~al.}(2016)\citenamefont
  {Mu\~noz}, \citenamefont {Kovetz}, \citenamefont {Dai},\ and\ \citenamefont
  {Kamionkowski}}]{Munoz:2016tmg}%
  \BibitemOpen
  \bibfield  {author} {\bibinfo {author} {\bibfnamefont {J.~B.}\ \bibnamefont
  {Mu\~noz}}, \bibinfo {author} {\bibfnamefont {E.~D.}\ \bibnamefont {Kovetz}},
  \bibinfo {author} {\bibfnamefont {L.}~\bibnamefont {Dai}},\ and\ \bibinfo
  {author} {\bibfnamefont {M.}~\bibnamefont {Kamionkowski}},\ }\href
  {https://doi.org/10.1103/PhysRevLett.117.091301} {\bibfield  {journal}
  {\bibinfo  {journal} {Phys. Rev. Lett.}\ }\textbf {\bibinfo {volume} {117}},\
  \bibinfo {pages} {091301} (\bibinfo {year} {2016})},\ \Eprint
  {https://arxiv.org/abs/1605.00008} {arXiv:1605.00008 [astro-ph.CO]}
  \BibitemShut {NoStop}%
\bibitem [{\citenamefont {Kader}\ \emph {et~al.}(2022)\citenamefont {Kader}
  \emph {et~al.}}]{Kader:2022jqp}%
  \BibitemOpen
  \bibfield  {author} {\bibinfo {author} {\bibfnamefont {Z.}~\bibnamefont
  {Kader}} \emph {et~al.},\ }\href@noop {} {\  (\bibinfo {year} {2022})},\
  \Eprint {https://arxiv.org/abs/2204.06014} {arXiv:2204.06014 [astro-ph.HE]}
  \BibitemShut {NoStop}%
\end{thebibliography}%

\end{document}